\documentclass[a4paper,fleqn,usenatbib]{mnras}

\usepackage{newtxtext,newtxmath}

\usepackage[T1]{fontenc}
\usepackage{ae,aecompl}
\usepackage[utf8]{inputenc}

\usepackage{graphicx}	
\usepackage{amsmath}	
\usepackage{xspace}

\usepackage{etoolbox}

\usepackage[dvipsnames]{xcolor}

\title[High resolution gas observations of CI Tau]{High resolution observations of molecular emission lines toward the CI Tau proto-planetary disc: planet-carved gaps or shadowing?}

\author[G. P. Rosotti et al.]{Giovanni P. Rosotti,$^{1}$\thanks{E-mail: rosotti@strw.leidenuniv.nl}, John D.  Ilee$^{2}$, Stefano Facchini$^{3}$, Marco Tazzari$^{4}$, \newauthor Richard A. Booth$^{4}$, Cathie Clarke$^{4}$, Mihkel Kama$^{5,6}$
\\
$^{1}$Leiden Observatory, Leiden University, P.O.~Box 9513, NL-2300~RA Leiden, the Netherlands\\
$^{2}$School of Physics and Astronomy, University of Leeds, Leeds LS2 9JT, UK\\
$^{3}$European Southern Observatory, Karl-Schwarzschild-Str. 2, 85748, Garching, Germany\\
$^{4}$Institute of Astronomy, University of Cambridge, Madingley Road, Cambridge CB3 OHA, UK\\
$^{5}$Department of Physics and Astronomy, University College London, Gower Street, London, WC1E 6BT, UK\\
$^{6}$Tartu Observatory, University of Tartu, 61602 Tõravere, Estonia
}

\date{Accepted XXX. Received YYY; in original form ZZZ}

\pubyear{2020}

\begin{document}
\label{firstpage}
\pagerange{\pageref{firstpage}--\pageref{lastpage}}
\maketitle

\begin{abstract}
Recent observations have revealed that most proto-planetary discs show a pattern of bright rings and dark gaps. However, most of the high-resolution observations have focused only on the continuum emission. In this Paper we present high-resolution ALMA band 7 (0.89mm) observations of the disc around the star CI Tau in the $^{12}$CO \& $^{13}$CO $J=3$--2 and CS $J=7$--6 emission lines. Our recent work demonstrated that the disc around CI Tau contains three gaps and rings in continuum emission, and we look for their counterparts in the gas emission. While we find no counterpart of the third gap and ring in $^{13}$CO, the disc has a gap in emission at the location of the second continuum ring (rather than gap). We demonstrate that this is mostly an artefact of the continuum subtraction, although a residual gap still remains after accounting for this effect. Through radiative transfer modelling we propose this is due to the inner disc shadowing the outer parts of the disc and making them colder. This raises a note of caution in mapping high-resolution gas emission lines observations to the gas surface density -- while possible, this needs to be done carefully. In contrast to $^{13}$CO, CS emission shows instead a ring morphology, most likely due to chemical effects. Finally, we note that $^{12}$CO is heavily absorbed by the foreground preventing any morphological study using this line.
\end{abstract}




\section{Introduction}

The Atacama Large Millimeter/Submillimeter Array (ALMA) is rapidly revolutionizing the field of proto-planetary discs thanks to transformational improvements in sensitivity and spatial resolution compared to the previous generation of sub-mm interferometers. One of the most important discoveries of the last few years is the realisation that most proto-planetary discs are not smooth, but have a wide variety of sub-structures such as gaps \citep{HLTau,Andrews2016}, spirals \citep{Perez2016,Boehler2018,Rosotti2020a} and crescents \citep{Casassus2013,vanderMarel2013,Cazzoletti2018}, at least for what concerns the dust emission.

While there is huge variety in disc structures, the results so far suggest that the most commonly found feature is azimuthally symmetric, colloquially called ``gaps and rings''. One of the best example is the DSHARP survey \citep{Andrews2018}, a high resolution (35-50 marcsec) continuum survey. Considering only the 18 discs in the survey around single stars, they all show azimuthally symmetric structures \citep{Huang2018rings}, with 3 also showing spirals and 2 showing crescents. Another example is a 120 marcsec survey in Taurus \citep{Long18}, which found rings in 12 discs out of 32. While lower resolution than DSHARP, the Taurus survey is less biased towards bright objects and therefore suggests that gaps and rings are common across the whole disc population. The discs in which no substructure was resolved are all compact (\citealt{Long2019}; see also \citealt{Facchini2019}), possibly indicating that the spatial resolution of the observations was not sufficient to find substructure, rather than a lack of substructure itself.

The vast majority of observations of substructure so far have focused only on the continuum. At the moment of writing, confirmed gaps using optically thin CO isotopologues as gas tracers have been observed only in HD169142 \citep{Fedele2017}, HD163296 \citep{Isella2016} and AS 209 \citep{Favre2019} (see also \citealt{Teague2017} for a gap in TW Hya using the CS molecule) at resolutions of 0.2-0.3", much lower than the 0.05" available in the continuum. This is easily understood since the continuum requires shorter integration times than line emission and therefore is more readily accessible by observations. However, it is well known that most of the proto-planetary disc mass is in the gas phase, and studying the continuum will therefore always offer a biased view of discs.

Opening a complementary view into the gas will be the task of the upcoming years. This is needed to answer the very questions opened by these continuum surveys. For example, the origin of the observed sub-structure is still unclear. Annular structures are naturally produced by planets \citep{Paardekooper2004,Dong2015,Rosotti2016} and could therefore be a powerful tool to study planet formation in action. On the other hand, other possibilities have been formulated, such as MHD effects \citep[e.g.,][]{Flock2015,Suriano2018,Riols2019} or opacity effects at snowlines \citep{Zhang2015,Okuzumi2016}. The latter option has been criticised in the last few years because in most cases it predicts that the rings should be at different spatial locations from where they are observed \citep{Huang2018rings,Long18}. However, a recent suggestion that snowlines can be thermally unstable (and therefore change location in the disc) may increase the viability of this explanation \citep{Owen2020}.  In the snowline interpretation, dust structures should be not accompanied by a similar change in the gas density and high-resolution gas observations can therefore rule out this possibility. In the planet case, deriving a planet mass only from the dust is extremely degenerate \citep{Zhang2018} and information on the gas can greatly reduce this degeneracy \citep{Facchini2018}.

Another question that can only be answered with a complementary view of gas in a disc is what is the potential for substructures to be the site(s) of planet formation?  By collecting large amounts of dust, the observed sub-structures are natural places where the planetesimal formation process could take place \citep{Eriksson2020} by triggering the streaming instability \citep{YoudinGoodman2005,Johansen2007}, instigating either a second-generation (if structures are created by planets) or first-generation (if other mechanisms are responsible for structures) round of planet formation (although this is not without its own challenges, see \citealt{Morbidelli2020}). Whether the conditions to trigger the instability are met, though, depends on how the dust behaves relative to the gas. This can be tested with gas observations \citep{Dullemond2018,Rosotti2020b}.

In this context, CI Tau is a source which shows three gaps and rings \citep{Clarke18,Long18}. It has a bright continuum disc \citep{Guilloteau2011} and several molecular species have been detected at sub-mm wavelengths \citep{Guilloteau2014,Guilloteau2016,Bergner2019,LeGal2019,Pegues2020}, making it a natural target for sensitive line studies. In this paper we report the results of high-resolution (100--150 marcsec) observations of this source in the $^{12}$CO \& $^{13}$CO $J=$3--2 and CS $J=7$--6 emission lines.

The paper is structured as follows. We first present the observations, data reduction and imaging parameters in section \ref{sec:observations}. We then present the continuum results in section \ref{sec:continuum} and the line results in section \ref{sec:results_lines}. In section \ref{sec:analysis_13co} we introduce radiative transfer models to interpret the $^{13}$CO observations and in section \ref{sec:discussion} we discuss our results in the wider context of the field. Finally we draw our conclusions in section \ref{sec:conclusions}.

\section{Observations and data reduction}
\label{sec:observations}

We obtained ALMA Cycle 5 DDT observations (Project ID: 2017.A.00014.S, PI: G. Rosotti) of CI Tau in band 7 on the 11th of December 2017 under very good weather conditions (with a mean precipitable water vapour column of $\sim$0.8 mm). Our target was observed with 43 antennas with baselines ranging from 15m to 3320m, and a total on-source integration time was 1h 18min. The correlator was set up to use four spectral windows, centred on 330.73GHz, 333.18GHz, 342.98GHz and 345.70GHz, respectively. The first spectral window was set to Time Division Mode (TDM) to observe the continuum with a bandwidth of 1.875GHz. The other spectral windows were set to Frequency Division Mode (FDM) with a spectral resolution of 564 kHz, corresponding to $\sim$ 0.5 km/s velocity resolution, to observe the $^{13}$CO $J=3$--2, CS $J=7$--6 and $^{12}$CO $J=3$--2 transitions. The bandwidth of each one of these three spectral windows was 937.5 MHz. 

To calibrate the visibilities we used the ALMA pipeline and the Common Astronomy Software Applications ({\tt\string CASA}, version 5.1.1). In addition to pipeline calibration, three rounds of phase-only self-calibration (with solution interval maximal for the first round, and then 360 and 180 seconds for the second and third round, respectively) were performed on the continuum data resulting in greater image fidelity in the outer disc, and improvement of the peak signal-to-noise by a factor of $\sim$2.0. Amplitude self calibration was attempted but it did not improve the signal-to-noise. These self-calibration solutions were then applied to the line data. In the paper we will analyse the resulting line data both with and without continuum subtraction. When continuum subtraction was applied, we performed it using the task {\tt\string UVCONTSUB} by fitting a first order polynomial to the line-free channels.

Continuum imaging was performed with the {\tt\string TCLEAN} task.  Using Briggs weighting with a robust parameter of 0.5, the resulting beam size for the dust continuum at a mean frequency of 338.2 GHz (886 $\mu$m; 0.89mm in the rest of the paper for simplicity) is 0.11 $\times$ 0.08$\arcsec$ (15.4 $\times$ 11.2 au assuming a distance of 140 pc\footnote{According to {\it Gaia} DR2 \citep{gaiadr2}, the distance to CI Tau is 158 pc. In this paper we adopt though 140pc because it was the distance assumed by \citet{Clarke18} and it simplifies the comparison with their results.}) with a position angle (PA) of 326$\degr$. We use the multi-scale deconvolve option with scales of 0, 6, 10 and 30 pixels, where a pixel is 0.016\arcsec.  We measure an rms noise level of 0.045 mJy/beam from emission free regions.

Line imaging was performed with a robust parameter of 1.0 and a channel spacing of 0.5\,km~s$^{-1}$.  The multi-scale deconvolve option was used with scales of 0, 5, 15 and 30 pixels (where a pixel is $0\farcs01$). We utilised Keplerian masks during the cleaning process (which are overlaid on the channel maps in Appendix \ref{sec:chans}) and cleaned to a threshold of $4\sigma$ (where $\sigma=2.98$ mJy/beam is the theoretical per channel sensitivity for observations made with the above settings under these conditions\footnote{\url{https://almascience.eso.org/proposing/sensitivity-calculator}}). These parameters were found to provide the best trade-off between spatial resolution and sensitivity, particularly in the outer regions of the disc.  The resulting final beam size was 0\farcs16$\times$0\farcs12 (${\rm PA} = 324.5^{\circ}$), with an rms noise level of 3.0 mJy/beam measured from line-free channels.

During imaging, we noticed that the clean beam was slightly non-Gaussian.  We applied the `residual scaling' method of \citet[][their Appendix A]{JvM_1995} in order to mitigate some of the effects of this on the final image quality.  Briefly, the process involves scaling the image residuals by a factor $\epsilon$, which is the ratio of the area of the clean beam (Gaussian) to the dirty beam (non-Gaussian).  These are then added to the model image to produce the image cube used for analysis.  Our measured values of $\epsilon$ were 0.32 for the $^{12}$CO, $^{13}$CO and CS image cubes, with final rms values of 1.1, 1.8, 1.2 mJy/beam, respectively.

\section{Observational results: continuum emission}
\label{sec:continuum}

\subsection{Band 7 (0.89mm)}

\begin{figure*}
\includegraphics[width=\columnwidth]{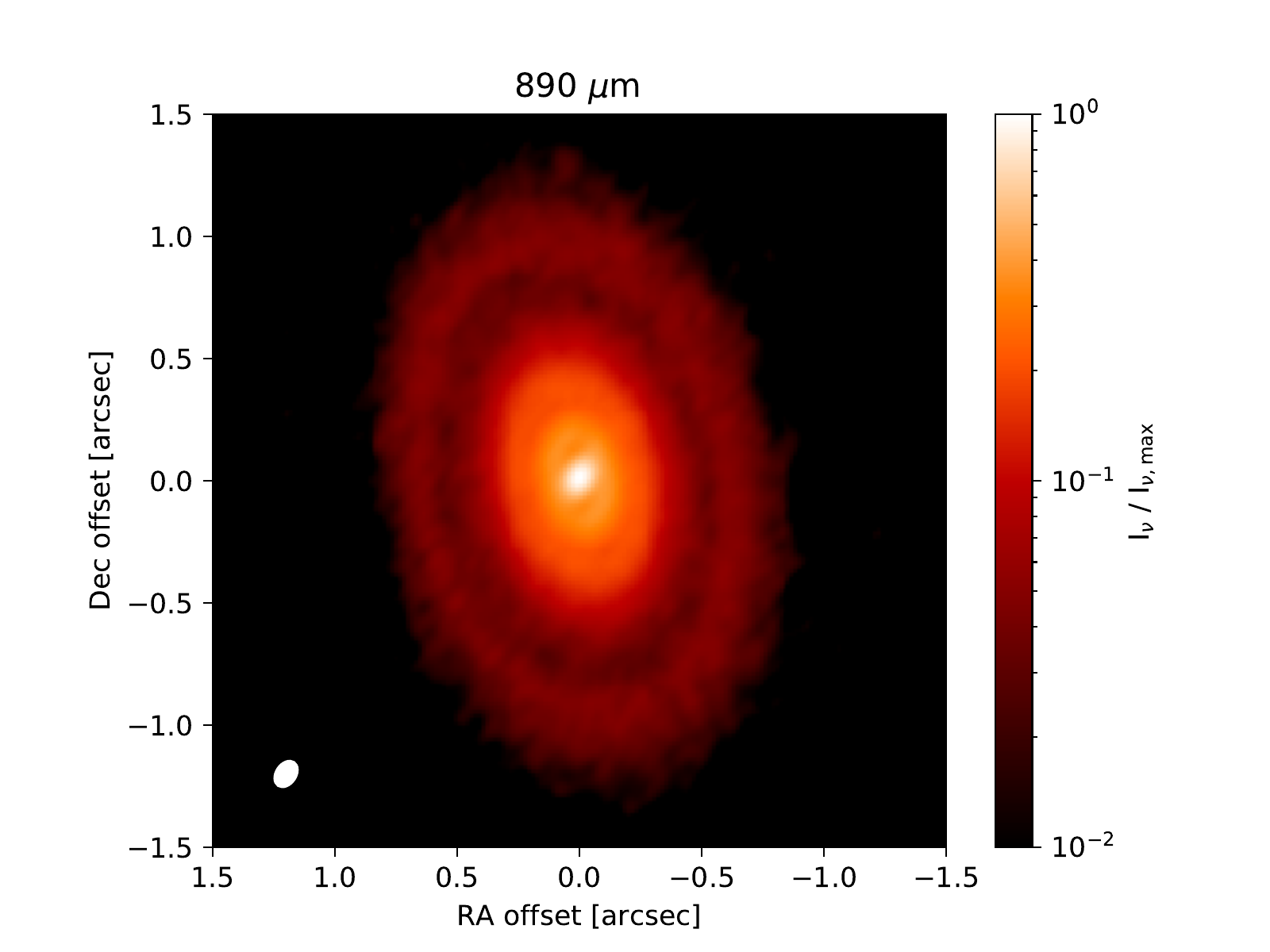}
\includegraphics[width=\columnwidth]{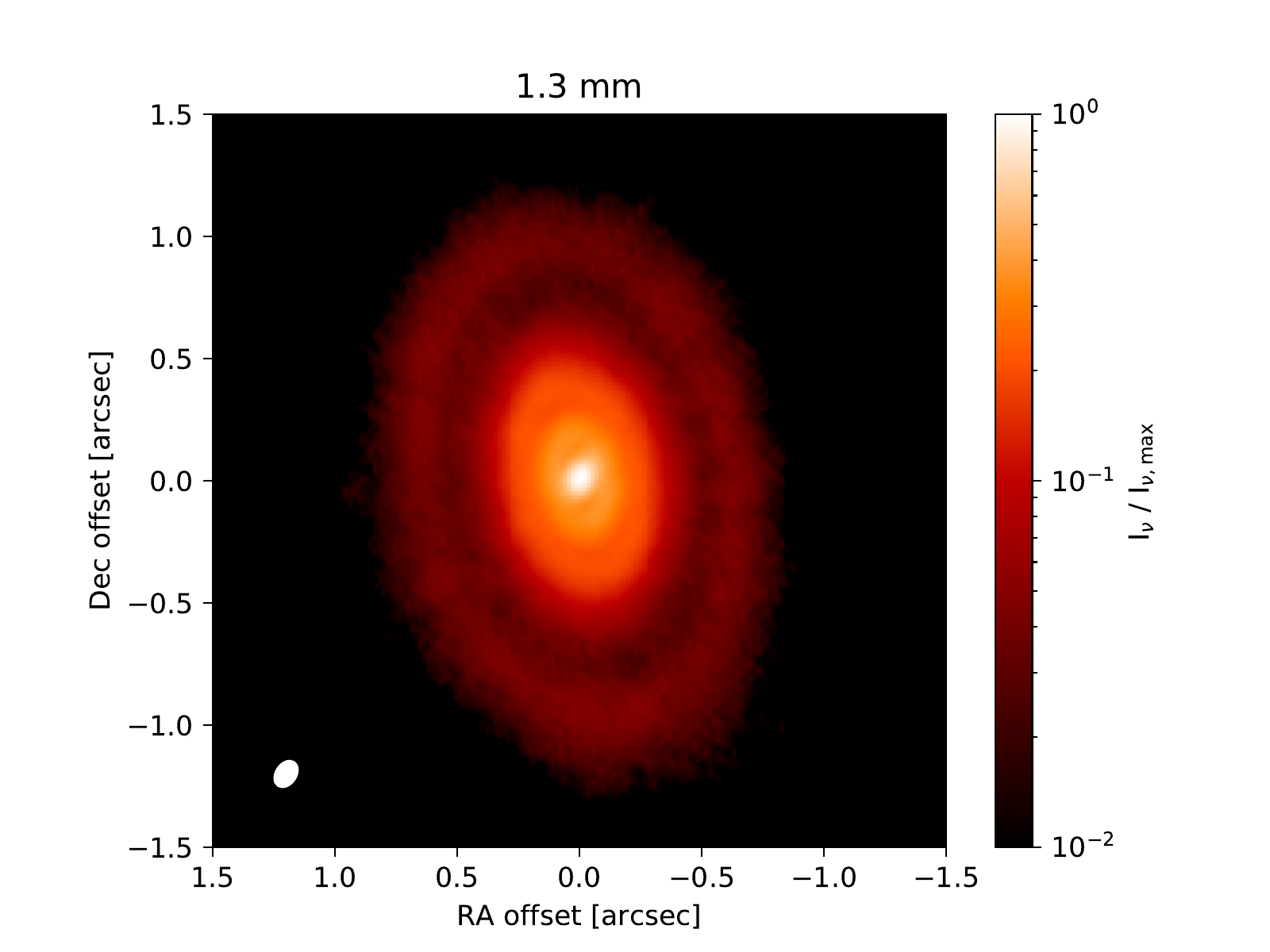}
\caption{\textbf{Left panel}: continuum image from the 0.89mm dataset we present in this paper. The emission is characterised by three dark, concentric rings. \textbf{Right panel}: continuum image from the dataset previously published by our team \citep{Clarke18} at 1.3mm degraded at the same resolution as the 0.89mm data.}
\label{fig:continuum_image}
\end{figure*}

\subsubsection{Image analysis}

We present the continuum emission at 0.89mm in the left panel of \autoref{fig:continuum_image}. The image is broadly similar to the one at 1.3mm previously published in \citet{Clarke18} and shows a series of three dark concentric gaps. At this resolution (a factor of $\sim$2.4 lower than \citealt{Clarke18}), the innermost gap is only barely resolved along the minor axis of the disc (which by a fortunate coincidence is almost aligned with the beam major axis), but still clearly present along the disc major axis. Recently, also \citet{Long18} presented 1.3mm observations of the source. Their resolution is 0.13 $\times$ 0.11$\arcsec$, slightly lower than what we present here. With their resolution, the innermost gap was not directly visible in the image, but they were still able to infer it through visibility modelling.

\begin{figure}
\includegraphics[width=\columnwidth]{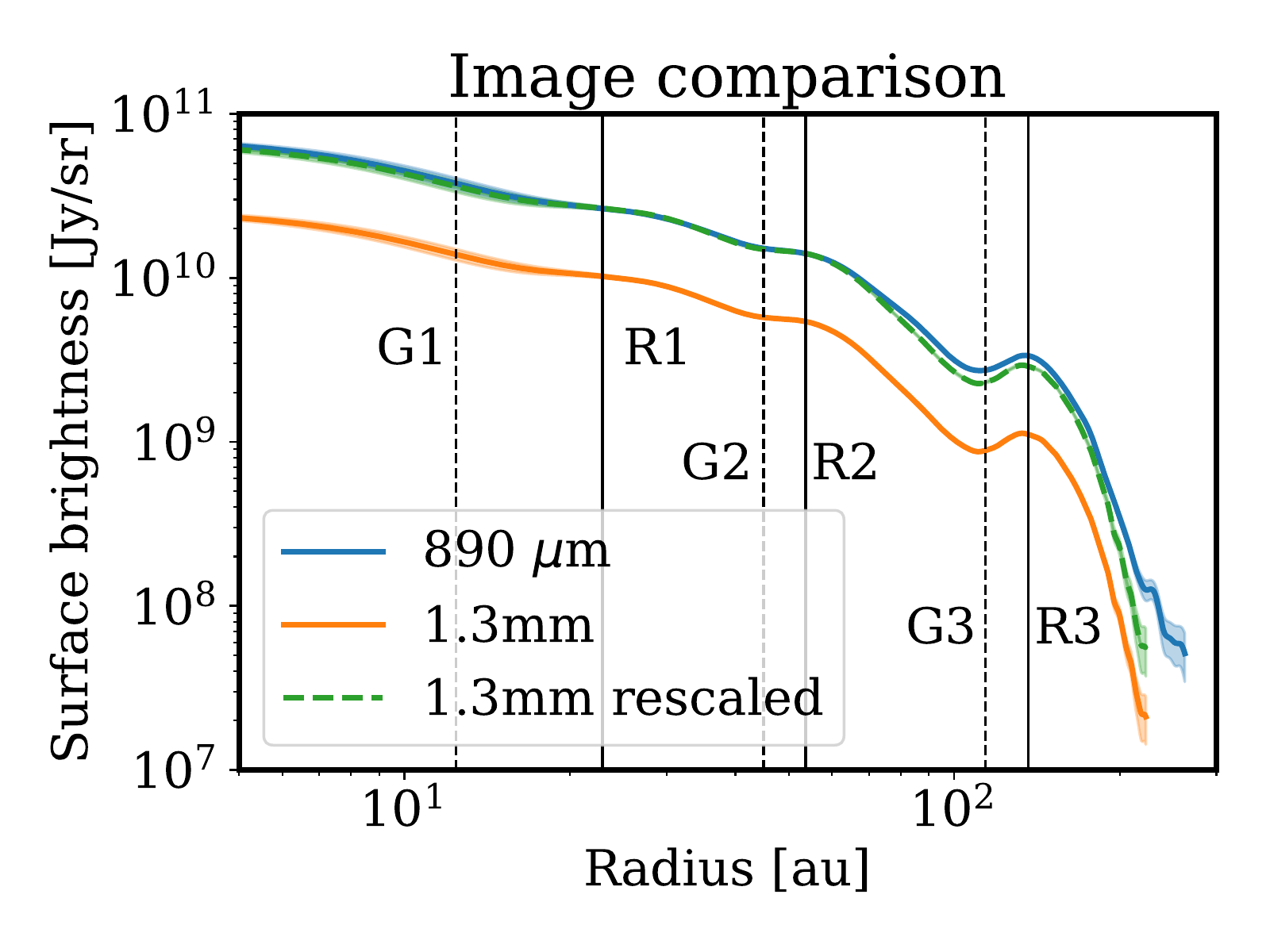}
\caption{Comparison between the deprojected radial profiles of the 0.89mm and 1.3mm continuum data. Dashed (solid) lines mark the central locations of the gaps (rings), which have been marked using the letter G(R). The shaded area represents the observational uncertainty; we plot the observational profiles only up to the point where the signal to noise ratio is 3. The two profiles are largely the same, but there is some difference; most notably, the emission profile decreases more steeply at 1.3mm than at 0.89mm in the outer parts of the disc.}
\label{fig:continuum_comparison_images}
\end{figure}

Qualitatively, when one considers the lower resolution of the 0.89mm dataset we present in this paper, the emission is largely similar to the 1.3mm data. This is confirmed by the right panel of \autoref{fig:continuum_image}, in which we plot the 1.3mm data degraded to the same resolution of the 0.89mm data. This has been accomplished using the {\tt \string restoringbeam} option in the {\tt \string tclean} task\footnote{We also experimented with an alternative method, consisting in using the {\tt \string uvtaper} option to achieve a similar beam and then the {\tt \string imsmooth} task to obtain exactly the same value, but we found no significant difference between the two methods.}.

To better quantify if there is any difference between the two images, we present in \autoref{fig:continuum_comparison_images} a comparison between the two de-projected radial profiles (using an inclination of 49\degr{} and a position angle of 101\degr{}, \citealt{Clarke18}). The shaded area around each line shows the 1-$\sigma$ uncertainty, quantified as the standard deviation along each (de-projected) circle, divided by the square root of the number of beams along the circle. We plot the profiles only up to the radius where the profile signal to noise ratio is 3. The two radial profiles show the three gaps and rings that we have previously described; for reference we have indicated them on the plot as the dashed and solid black vertical lines, respectively. Their locations are 12, 45 and 114 au for the gaps, marked as G1, G2 and G3 on the plot, and 23, 54 and 136 au for the rings marked as R1, R2 and R3 on the plot. Finally, for clarity, in the figure we also show the 1.3mm continuum profile multiplied by an arbitrary factor, to better illustrate the differences from the 0.89mm data.
The figure shows that, while broadly the same, there are in fact some difference between the two images. In particular, the emission profile at 0.89mm appears to have a slightly steeper drop in the outer part of the disc (i.e., outside the second gap, beyond $\sim$ 70 au). We will analyse more in detail these differences in section \ref{sec:spectral_index}.

\subsubsection{Visibility modelling}

In order to better characterise the emission, we fit the continuum visibilities with an axisymmetric parametric model consisting of an envelope and three gaps. The analysis largely follows \citet{Clarke18} and we refer the reader to that work for a more extensive discussion. In summary, we describe the emission as the superposition of a background, represented by an exponentially tapered power-law, and three gaps, that we describe using logistic functions (a functional form chosen for its flexibility). The envelope is described by 5 free parameters, while each gap is described by 6. Since we also fit for the disc inclination, position angle and offset from the phase center, we have a grand total of 27 free parameters.

We fit the parameters of the model using the Bayesian Markov-Chain sampler EMCEE \citep{EmceePaper}, which enables us to estimate both the best fit and the uncertainties on the parameters. For each model realisation, we use the code \textsc{GALARIO} \citep{GalarioPaper} to compute synthetic visibilities and compare them to the measurements.

\begin{figure}
\includegraphics[width=\columnwidth]{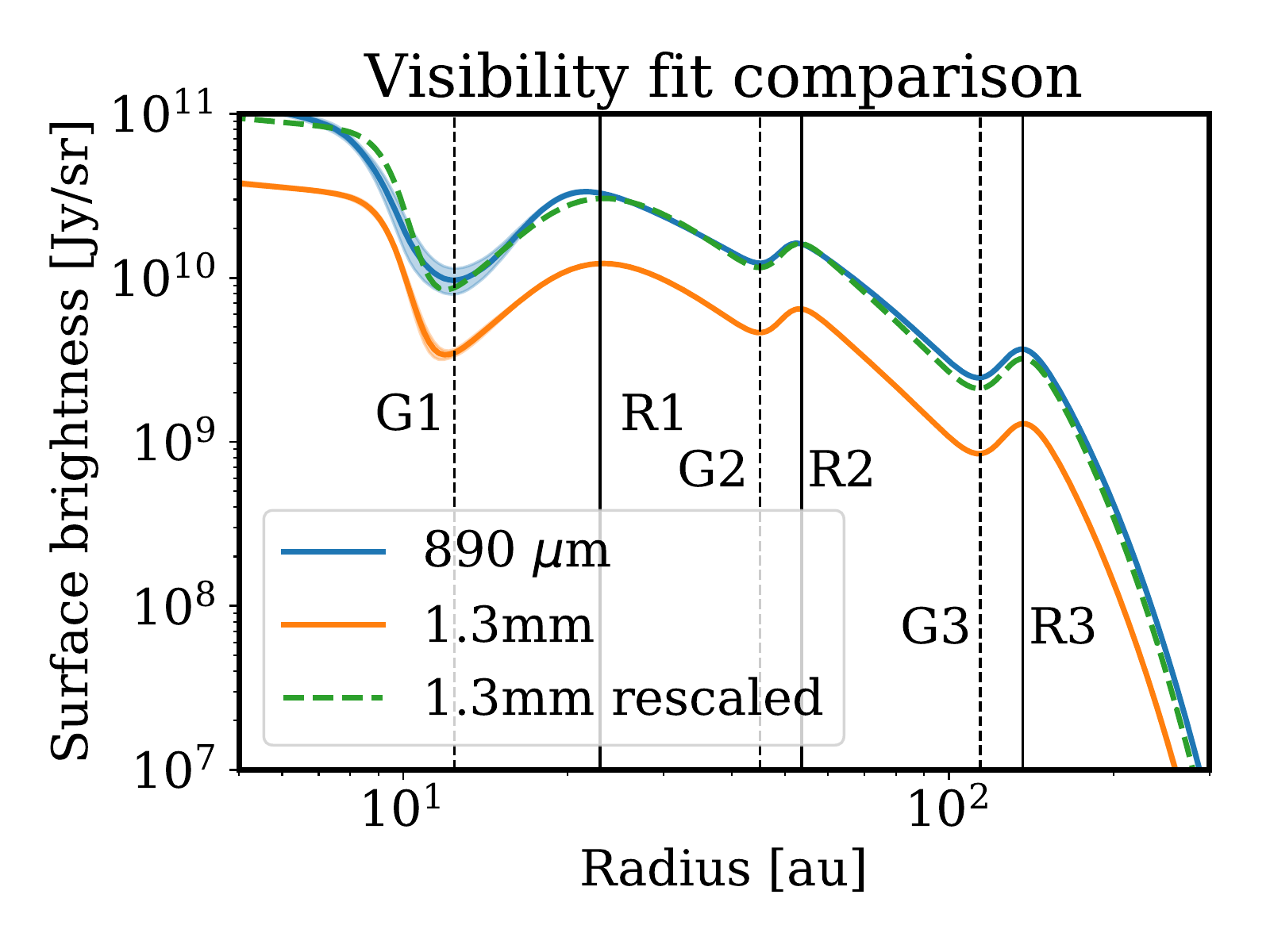}
\caption{Comparison between the two \textsc{GALARIO} fits to the visibility data at 0.89 and 1.3mm. The shaded area represents the uncertainty in the fit. The result of the fit confirms the steeper decrease in emission in the outer parts of the disc.}
\label{fig:continuum_comparison_uvfit}
\end{figure}

We show in \autoref{fig:continuum_comparison_uvfit} the comparison between the resulting best fit profiles to the 1.3mm \citep{Clarke18} and 0.89mm (this work) data. The shaded area around each line illustrates the statistical uncertainty in the best fit and is the standard deviation computed from 1000 random draws from the chains in the sampler. At the resolution of the dataset we present in this work, the fit is not able to constrain well the properties of the innermost gap G1, which is reflected in the larger uncertainty around the best fit when compared to the 1.3mm data. That being said, considering the difference in resolution between the two datasets (a factor of $\sim$2.4) and that the innermost gap G1 is barely resolved in the image plane, the visibility modelling still performs very well. Outside the first gap, the statistical uncertainty becomes very small and it is likely that the true uncertainty is dominated by the systematic uncertainty connected with chosen functional form for the parametric fitting.

In \autoref{fig:continuum_comparison_uvfit}, we have also plotted the 1.3mm fit rescaled by an arbitrary factor to better illustrate the differences between the two wavelengths. The comparison confirms the difference between the two bands already noted by the comparison between the two images (\autoref{fig:continuum_comparison_images}).

\subsection{The spectral index between 1.3 mm and 0.89 mm}
\label{sec:spectral_index}

\begin{figure}
    \centering
    \includegraphics[width=\columnwidth]{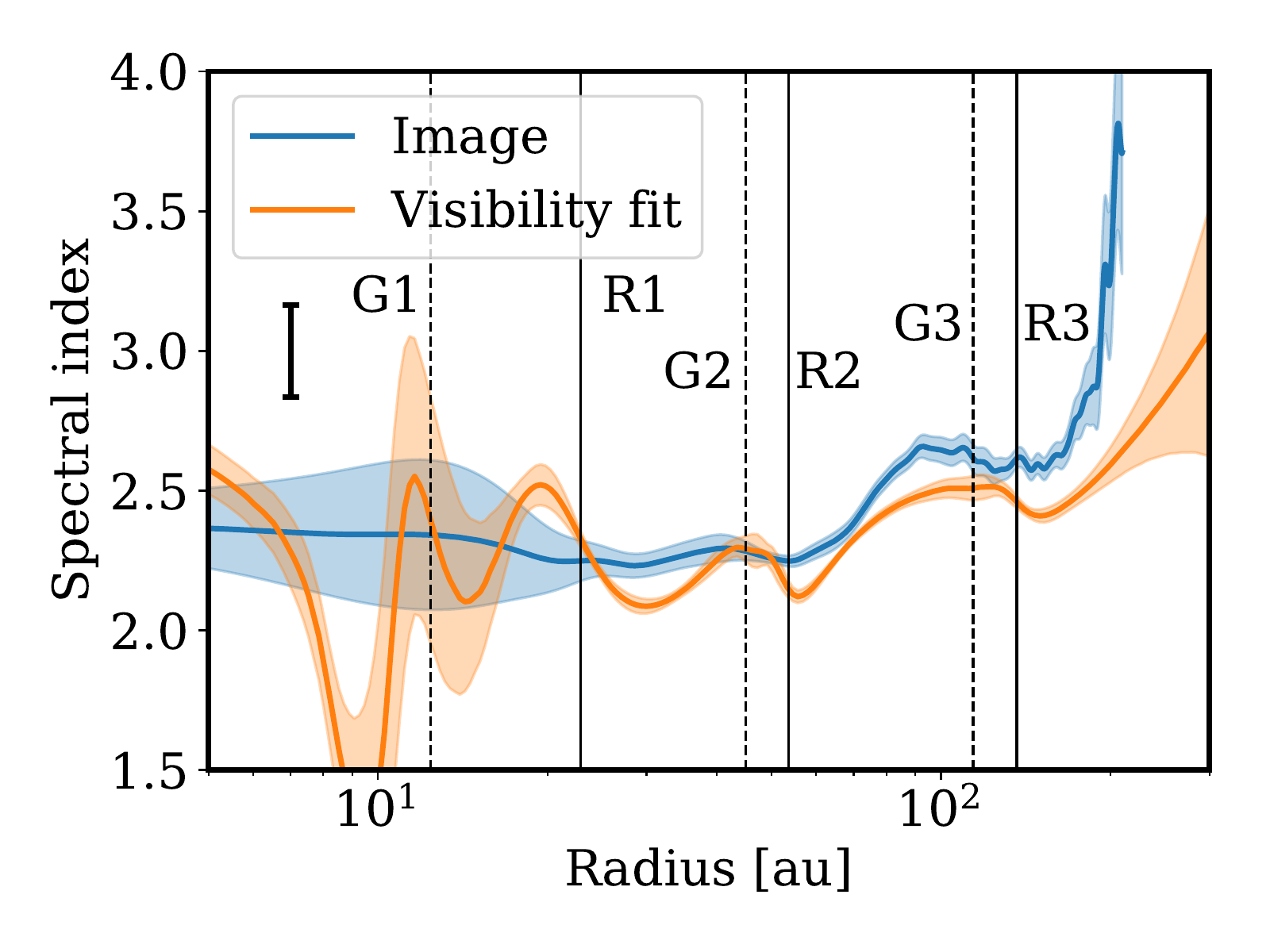}
    \caption{Spectral index computed between the 1.3 and 0.89mm continuum data using both the images and the fits to the visibilities. We have marked the size of the systematic error in the spectral index coming from absolute flux calibration with the black bar. After the second ring, the spectral index increases with radius, in line with the steeper decrease of the emission observed when comparing the 1.3 and 0.89mm brightness profiles. Variations between gaps and rings are hard to assess due to the limited spatial resolution.}
    \label{fig:continuum_spectral_index}
\end{figure}

Having constrained the deprojected radial profiles of the emission from the analysis of the images and the visibilities, we now proceed to compute the spectral index $\alpha=\mathrm{d}\log{F_\nu}/\mathrm{d}\log{\nu}$ between the two wavelengths. The spectral index is a useful and widely studied quantity since it provides information about grain size and optical depth \citep[e.g.,][]{Testi2014}. While in the pre-ALMA era it was only possible to study spatially integrated spectral indices \citep{2005ApJ...631.1134A,Ricci2010a,Ricci2010b}, with few exceptions \citep{Perez2012,Tazzari2016}, the study of spatially resolved spectral indices is now becoming routinely feasible both around single objects \citep{Tsukagoshi2016,Huang2018TwHya,Dent2019,Macias2019,2020ApJ...891...48H,Long2020,Tazzari2020} and for large samples of discs \citep{2020arXiv201002249T}, though the presence of (potentially optically-thick) sub-structures and the limited angular resolution significantly complicates the interpretation of these results.

We plot the spectral index in \autoref{fig:continuum_spectral_index} for the two methods. Assuming a 10 percent flux calibration error at both wavelengths, the spectral index has an absolute systematic uncertainty of 0.33, which we mark on the plot with the black error bar. This does not affect however the shape of the spectral index variation with radius. The shaded area represents the statistical uncertainty coming from the limited signal to noise of the observations. To make sure we do not introduce artefacts, for the image plane analysis we have plotted the spectral index only up to the radius where the signal to noise ratio is at least 3 for both wavelengths. 

The spectral index is almost constant up to the second ring R2 and then increases with radius, confirming that the decrease of the emission towards the outer radii is steeper at 1.3mm than at 0.89mm. However, the increase is not monotonic: both methods find a local decrease in the spectral index at the location of the third ring R3, which could be due to either larger grains or to the increased optical depth. For what concerns the first gap G1 and ring R1, the large oscillations in the inner part of the disc imply that at this resolution these datasets cannot be used to study the spectral index at these spatial scales, with the resolution at 0.89mm the limiting factor. The large oscillations in the visibility fit at these distances are likely a result of the inflexibility resulting from a chosen functional form for the parametric fit. For the second gap G2 and ring R2, the visibility fit indicates an increase in the spectral index in the second gap G2 and a decrease in the second ring R2, which again can be interpreted either as a variation of grain size or increased optical depth. Instead, the images do not show a variation in the spectral index for the second gap/ring, due to a lack of spatial resolution. 

These results highlight that these two datasets cannot be used to reliably study the spectral index variation in the gaps and rings, mostly due to the limited spatial resolution of the 0.89mm data. On the other hand, the increase of the spectral index in the outer part of the disc is robust. Because emission is likely optically thin at these radii, differently from the rings, this suggests a decrease in grain size towards the outer parts of the disc. In section \ref{sec:radial_extent} we will show that the disc is larger in gas emission than in the continuum. The increase in spectral index therefore suggests that the extent of the continuum disc is only tracing the extent of millimetre particles rather than the true extent of the disc. However, given the large error bar coming from the absolute flux calibration, we will not attempt in this paper a more quantitative analysis of the spectral index. This requires the inclusion of longer wavelength to provide more leverage and reduce the systematic uncertainty, for example the 2mm and 3mm bands available with ALMA or even longer wavelengths that can be accessed only with the VLA \citep{2016A&A...588A.112G,2019ApJ...883...71C}. We defer such an analysis to future studies when the relevant data will be available.

\section{Observational results: line emission}
\label{sec:results_lines}

Many central channels of $^{12}$CO are severely affected by foreground absorption (see \autoref{fig:12CO_observations}), so that only part of the emission from the disc can be recovered.  Due to this, we do not perform a detailed analysis of the $^{12}$CO emission, and instead concentrate our efforts on $^{13}$CO and CS.  We first outline how we compute emission maps and radial profiles from the cubes, and then discuss the two emission lines.

\begin{figure*}
\centering
\includegraphics[width=\textwidth]{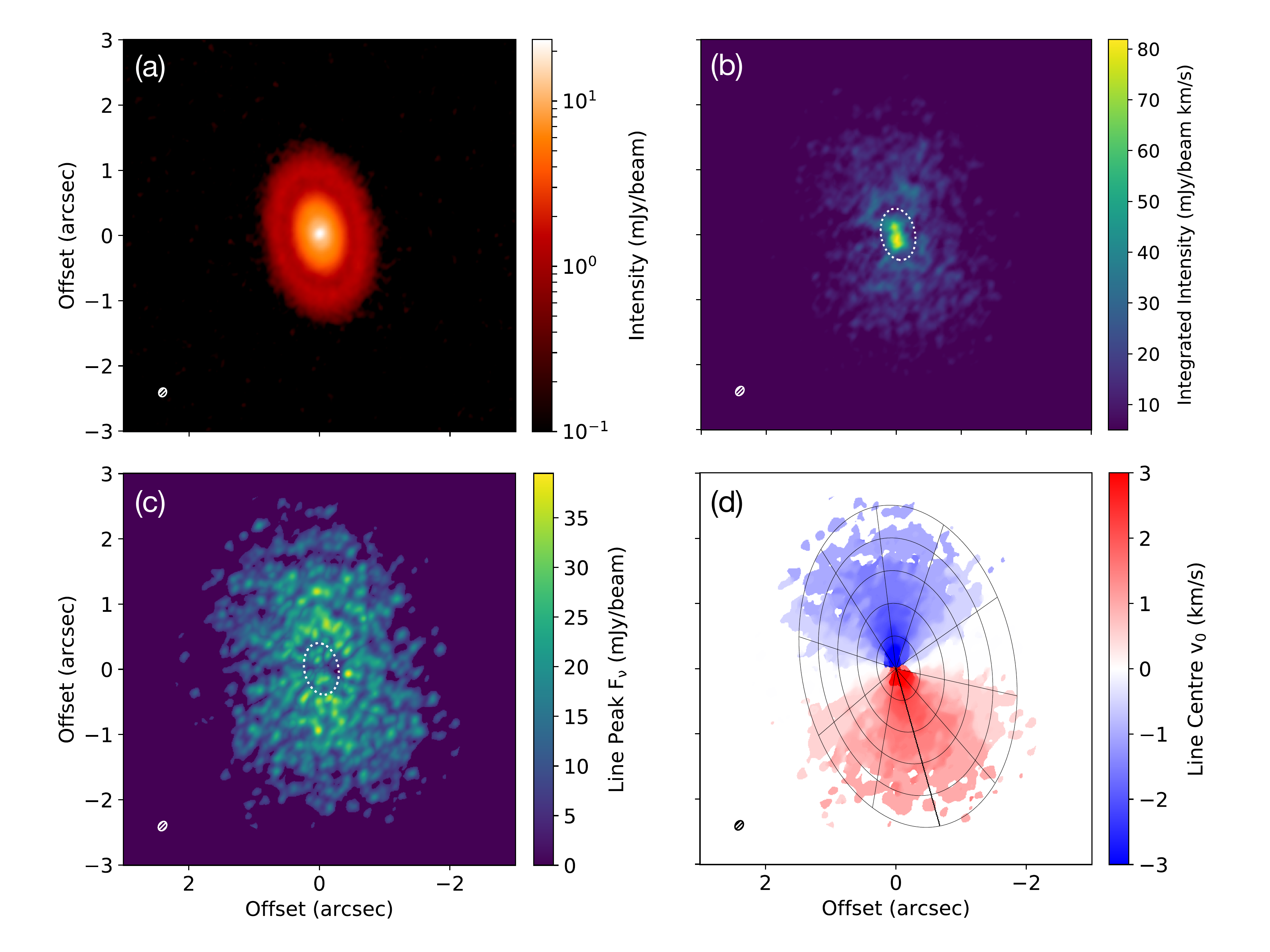}
\caption{Maps of $^{13}$CO (3--2) emission toward CI Tau -- a) 0.89mm continuum emission, b) integrated line intensity, c) line peak and d) line centre. The white ellipse shows the location of the gap we discuss in the text. The line centre panel is overlaid with a conical surface of aspect ratio $z/r=0.1$.}
\label{fig:13CO_observations}
\end{figure*}

\begin{figure*}
\centering
\includegraphics[width=\textwidth]{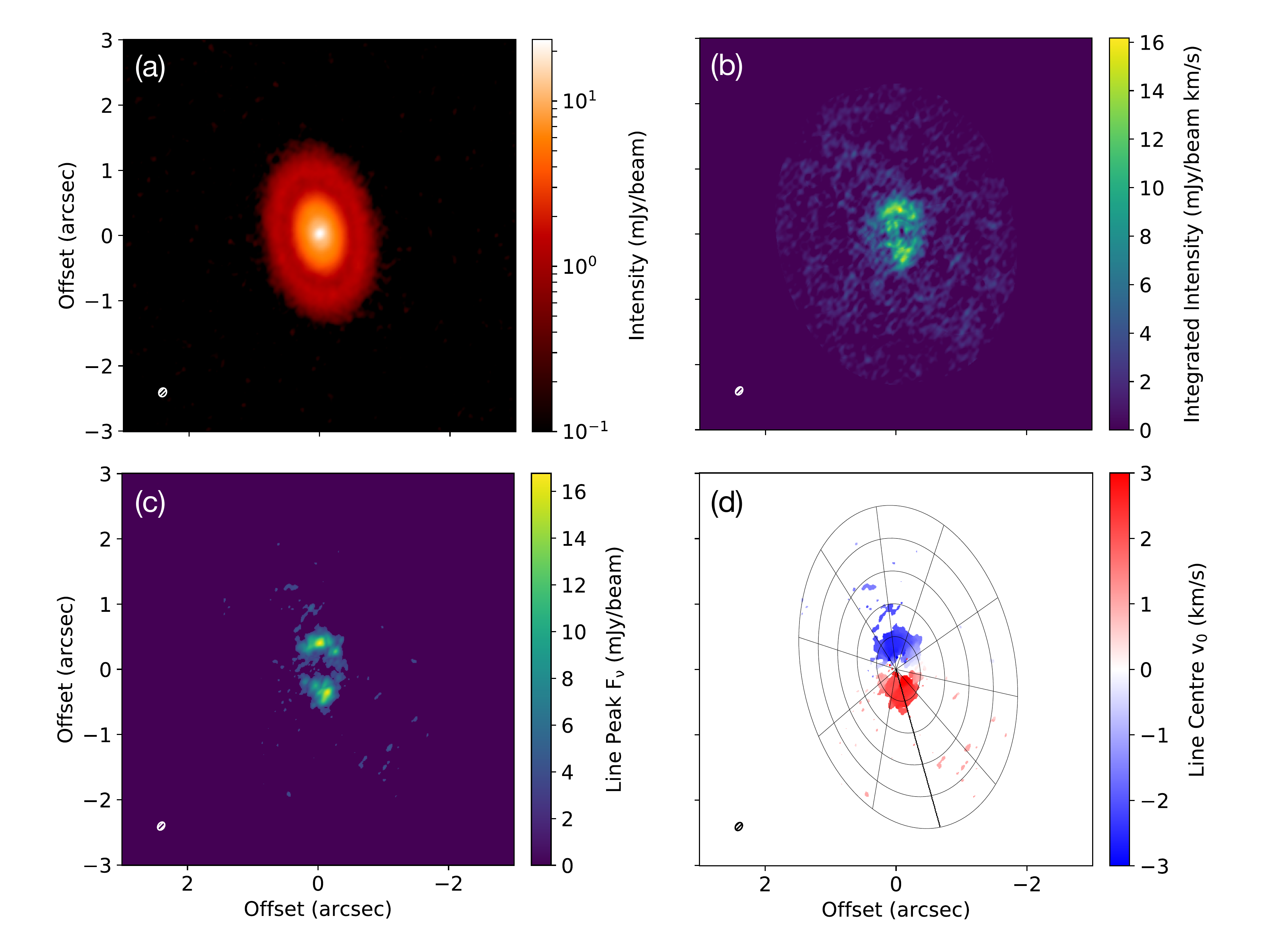}
\caption{As \autoref{fig:13CO_observations}, but for CS (7--6). The line centre panel is overlaid with a conical surface of aspect ratio $z/r=0.1$.}
\label{fig:CS_observations}
\end{figure*}

\begin{figure*}
\centering
\includegraphics[width=0.49\textwidth]{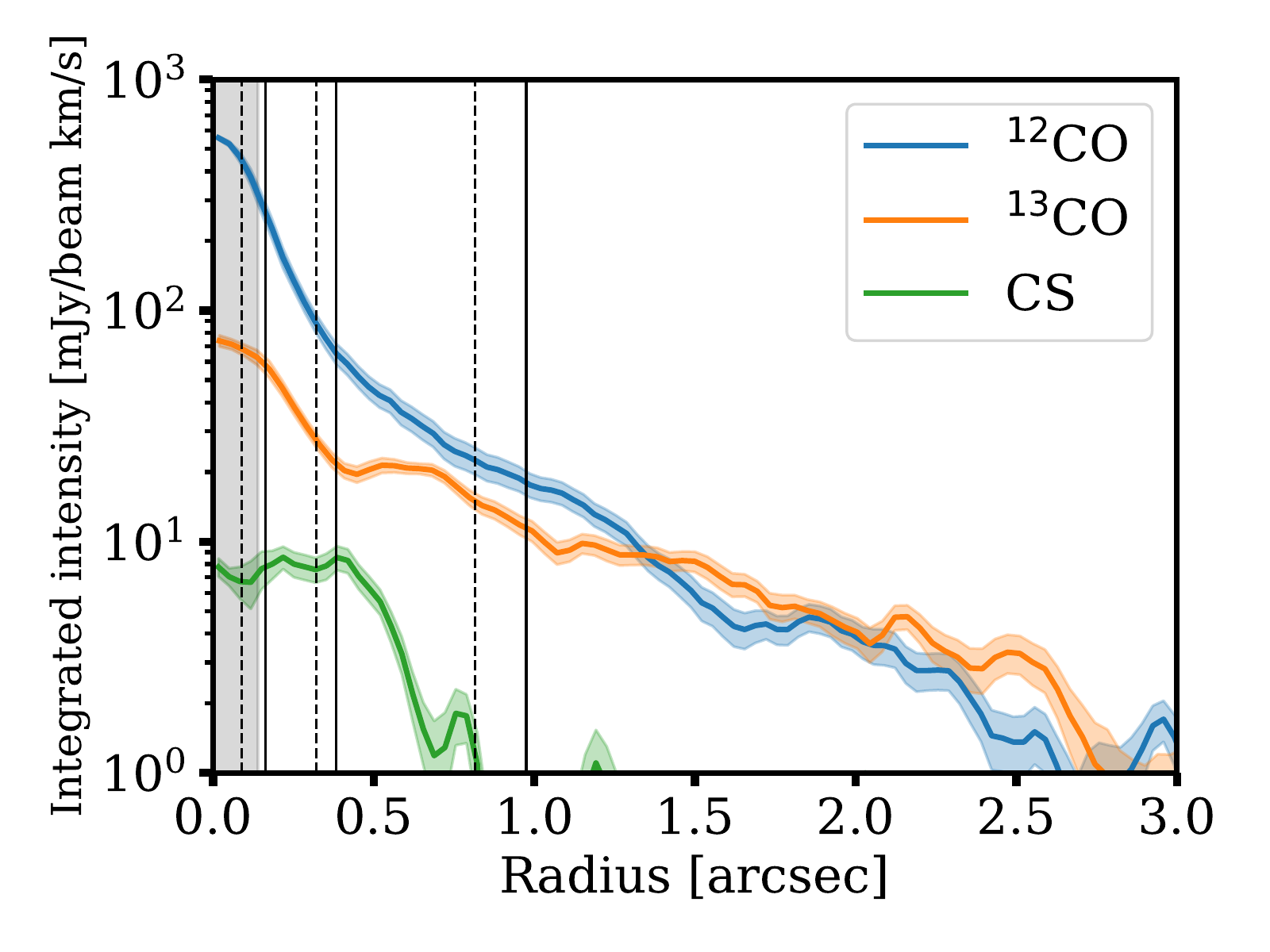}
\includegraphics[width=0.49\textwidth]{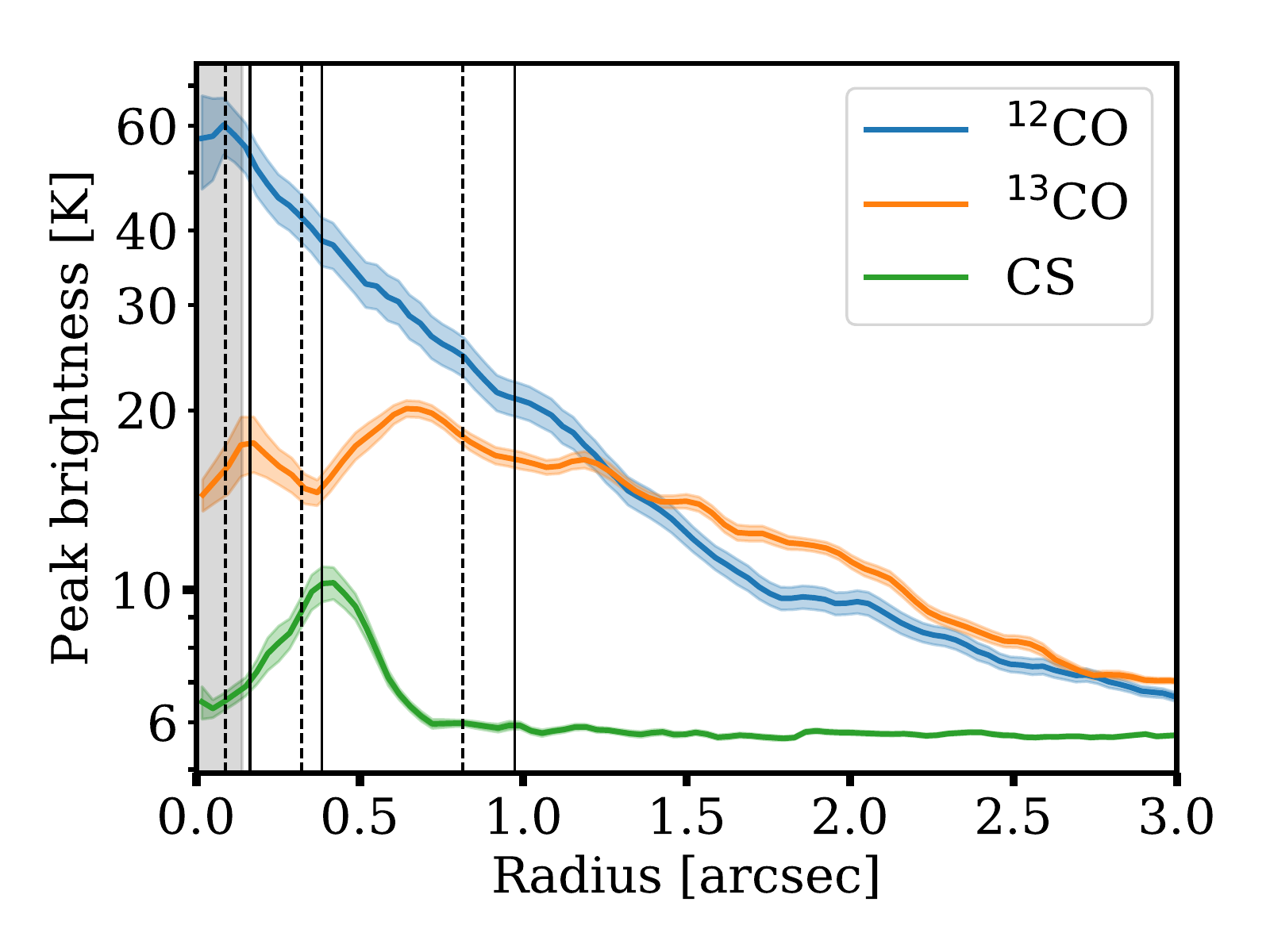}
\caption{Radial profiles for the line emission (computed using the {\tt\string gofish} package, \citealt{teague_gf_2019}). \textbf{Left panel}: integrated intensity (moment 0) map. \textbf{Right panel}: peak brightness map. The vertical lines represent the location of the continuum features, gaps (dashed) and rings (solid), respectively. The shaded grey area represents the size of the beam (average FWHM of 0.14 \arcsec).}
\label{fig:radial_profiles}
\end{figure*}

\subsection{Analysis of line cubes}

For each line cube, we use the {\tt\string bettermoments} \citep{teague_bm_2018} package to compute a series of spectral moments (see Figures \ref{fig:12CO_observations}, \ref{fig:13CO_observations}, \ref{fig:CS_observations}).  In the first instance, these are the integrated intensity (zeroth moment) and intensity weighted velocity (first moment) assuming a clip of 3$\sigma$ for the latter.  We also collapse each cube using the `quadratic' method, which results in maps of line peak $F_{\nu}$ and line centre $v_{0}$.  These quantities are analogous to the traditional eight and ninth spectral moments, but have the added advantage of sub-channel velocity precision and reducing the effect of noise, significantly improving the fidelity of the resulting maps \citep[see][for full details]{teague_bm_2018}. During the calculation of all of the above maps, we make use of the Keplerian masks shown in Figure \ref{fig:channel_maps}.

The calculation of an azimuthally-averaged radial profile is complicated by the fact that the line emission can originate from a surface inclined above the midplane of the disc.  This effect is particularly noticeable in the $^{12}$CO (3--2) line peak and centre maps.  We therefore utilise the {\tt\string gofish} package \citep{teague_gf_2019} in order to overlay a conical emission surface on each of the intensity maps, and use this surface to perform the azimuthal averaging. The aspect ratios ($z/r$) of the emission surfaces found to best follow the line emission were 0.3, 0.1 and 0.1 for the $^{12}$CO, $^{13}$CO and CS, respectively\footnote{Attempts at more rigorous fits with the {\tt\string eddy} code \citep{2019JOSS....4.1220T} did not converge. We stress that the precise value of $z/r$ does not have a significant impact on the recovered radial profiles using {\tt\string gofish}.}. These surfaces are overlaid on the line centre panels in Figures \ref{fig:12CO_observations}, \ref{fig:13CO_observations},\ref{fig:CS_observations}. Radial profiles are then computed using this surface, with uncertainties calculated as the standard deviation scaled by the number of beams per annulus to account for correlated noise.  These profiles are shown in Figure \ref{fig:radial_profiles}.

\subsection{$^{13}$CO $J=3$--2}
\label{sec:13co}

\autoref{fig:13CO_observations} shows the continuum-subtracted moment maps for the $^{13}$CO emission. We can see that in this case we recover emission from the whole disc, implying that the column of material absorbing the $^{12}$CO emission is too little to significantly absorb $^{13}$CO. Given that the ratio of the isotopic abundance between the two species is large, $\sim 70$, this is a condition relatively easy to satisfy.

Inspection of the projected velocity map shows that in this case the emission is coming from a surface closer to the midplane than for $^{12}$CO. On the other hand, the intensity map (both integrated and peak intensity) shows that the disc is larger in gas emission than it is in the continuum (note the different scale from \autoref{fig:continuum_image}), clearly extending beyond 2\arcsec{}, whereas the continuum has a steep drop beyond 1.5\arcsec{}. In terms of morphology, the maps do not show very conspicuous features (in contrast the continuum image), but there is a hint of a gap around 0.4\arcsec{} that we marked with a white, dashed ellipse.

To better quantify the morphology of the images, we have deprojected the two maps and studied their radial profiles, which we show in \autoref{fig:radial_profiles}. Gas emission is detected up to 2.5\arcsec{}, after which there seems to be a drop-off in the emission profile, although we caution that the S/N in the outer parts of the disc is limited. To compare quantitatively the disc size in gas and continuum, we have estimated the 68 per cent flux radius in both cases. For the gas, the 68 per cent radius is 1.8\arcsec{}. Conversely, for the continuum we obtain a value of 0.8\arcsec{}, yielding a ratio of $\sim$2.25. Given that our observations have a nominal maximum recoverable scale of 1.3\arcsec{}, our estimate of the gas radius is most likely a lower limit and the emission could be even more extended. It is well known that discs are in general larger in CO line emission than in the continuum; we shall go back to the implications of this in section \ref{sec:radial_extent}.

The radial profile also confirms that there is indeed a gap in the $^{13}$CO emission at a radius of 0.4\arcsec{}, which can be seen both in the integrated intensity and in the peak brightness maps. To make the comparison with the continuum structure easier, we have overplotted as vertical lines the location of the structures found in the continuum. Surprisingly, the dip in emission in the line is not centred at the location of a continuum gap, but at the location of a continuum ring. We will discuss the physical interpretation of this structure in section \ref{sec:gap_13co}.

\subsection{CS $J=7$--6}
\autoref{fig:CS_observations} shows the continuum-subtracted moment maps for the CS emission. Also in this case we do not see evidence of foreground absorption and the projected velocity map shows that the emission is coming from a surface closer to the midplane than for $^{12}$CO.

The morphology of the emission is very different from $^{13}$CO. It is clear that the emission does not peak at the center of the disc (contrast this with the $^{13}$CO integrated intensity map). Instead, the emission has a ring morphology, as confirmed by both the integrated intensity and peak brightness maps. Like for $^{13}$CO, we show in \autoref{fig:radial_profiles} the deprojected radial profiles. These confirm that the emission has a local maximum at $\sim$0.5\arcsec{}.

\section{Analysis of the $^{13}$CO emission}
\label{sec:analysis_13co}

In this section we set up radiative transfer models that we use to analyse and interpret the $^{13}$CO emission. We first describe the general setup of the models and then discuss whether or not we should perform continuum subtraction on the datacubes, using a few models as illustrative example. We then analyse the disc extent and the $^{13}$CO gap in comparison with the models.

\subsection{Radiative transfer general setup}
\label{sec:radiative_transfer}

\begin{table}
\centering
\caption{Dust and gas surface density for the radiative transfer models we present in section \ref{sec:analysis_13co}.}
\label{tab:rt_sigma}
\begin{tabular}{p{15mm}p{25mm}p{25mm}}
Model & Dust surface density & Gas surface density \\
\hline \hline
Constant dust-to-gas ratio & Continuum profile of \citet{Clarke18} & $\Sigma_\mathrm{dust}$ re-scaled by 100\\\hline
Hydro & Hydro simulation of \citet{Clarke18} & Hydro simulation of \citet{Clarke18}\\\hline
Hydro + ring & Hydro simulation of \citet{Clarke18} + ring at 20 au& Hydro simulation of \citet{Clarke18}\\
\end{tabular}
\end{table}

We use the code RADMC-3D\footnote{\url{http://www.ita.uni-heidelberg.de/~dullemond/software/radmc-3d/}} to run radiative transfer models of the source. In the radiative transfer calculation we use a spherical mesh with $N_r$=330 and $N_\theta$=80 grid cells in the radial and poloidal direction, respectively. Because the disc does not show deviations from axisimmetry, we only consider axisymmetric models. The grid extent is [1,300] au for the radial grid and [0,$\pi/2$] for the poloidal. Dust is implemented using two separate populations, ``small'' and ``large'' grains. Small grains are always assumed to have a constant dust-to-gas ratio, which we take to be 0.1 percent. The surface density of large grains and of the gas is prescribed as we will discuss later for each specific model, but see \autoref{tab:rt_sigma} for a summary. Given a surface density, we distribute the material vertically assuming that the density $\rho$ follows a Gaussian function $\rho(z) \propto \exp(-z^2/2H^2)$. We prescribe the initial scale-height $H$ as $H(r)=0.1 (r/100\  \mathrm{au})^{0.15}$. We assume that large grains are settled to the midplane and take their scale-height to be a factor of 5 smaller than that of the gas. We compute opacities as in \citet{Tazzari2016} following models by \citet{Natta:2004yu} and \citet{Natta:2007ye}, using the Mie theory for compact spherical grains with a simplified version of the volume fractional abundances in \citet{1994ApJ...421..615P}, assuming a composition of 10\% silicates, 30\% refractory organics, and 60\% water ice. We assume that for each grain population the grain size distribution is a power-law $n(a)\propto a^{-q}$ for $a_\mathrm{min}\leq a\leq a_\mathrm{max} $ with an exponent $q=3$. ``Small'' grains have $a_\mathrm{max}=1 \ \mu$m, while ``large'' grains have $a_\mathrm{max}= 1$ mm.

We first perform a thermal Monte Carlo run using $50 \times 10^6$ photons to compute the dust temperature. We then use this temperature to update the disc scale-height and recompute the temperature accordingly. Tests showed that already after one iteration we do not see a significant change in the $^{13}$CO emission. Once the temperature is known, we can populate the disc of CO. We parametrize the behaviour of CO largely following \citet{WilliamsBest2014}. Firstly, we assume that the gas temperature is the same as the dust temperature. We assume that CO is frozen out onto the grains when the temperature is smaller than 19K and we assume it is dissociated when the vertical column is smaller than $N_\mathrm{CO} < 10^{15} \ \mathrm{cm}^{-2}$ \citep{vanDishoeckBlack1988,Visser2009}, which well reproduces the results of thermo-chemical models \citep[e.g.,][]{Trapman2019}. Where CO is neither frozen out nor dissociated, we assume an abundance of $10^{-5}$. We do this rather than using the standard abundance of $10^{-4}$ because tests showed that using the surface density profile of the hydrodynamic simulation presented in \citet{Clarke18} led to an overestimate of the $^{13}$CO emission, implying that the CO column had to be reduced. We note that in our setup CO abundance and total gas surface density are degenerate - the same result could have also be obtained by reducing the total gas mass. Considering a ratio of isotopic abundance of 70 between $^{12}$C and $^{13}$C, we use an abundance of $^{13}$CO of 1.5 $\times 10^{-7}$. As a last step, we produce a synthetic datacube assuming Local Thermodynamic Equilibrium; we use the Leiden Molecular Database\footnote{\url{https://home.strw.leidenuniv.nl/~moldata/}} \citep{Schoeier2005} to set the frequency and Einstein coefficients of the $J=3$--2 transition. We assume an inclination of 49\degr{} and a position angle of 101\degr{} \citep{Clarke18} to generate the cube. Finally the synthetic datacube is convolved with a Gaussian beam (tests with a full CASA simulation did not reveal significant differences) matching the beam of the observations (0.16 $\times$ 0.12\arcsec{}). 

\subsection{On sub-structure in continuum subtracted maps}
\label{ssec:cont_sub}

As we already highlighted, the location of the dip in $^{13}$CO does not coincide with the location of a continuum \textit{gap}, but rather the location of a continuum \textit{ring}. It is difficult to imagine a scenario in which the gas would be depleted at the same location where the dust accumulates. This begs the question of whether the observed dip can be simply interpreted as due to a reduction in the gas surface density or if it has a different origin.

Until now we have followed the standard practice of performing continuum subtraction to analyse line datacubes. We argue that this practise is not justified in this case and we show through an example how it can lead to spurious sub-structures in emission line maps. We provide a detailed explanation of why this is the case in appendix \ref{append:continuum_subtraction}.

\begin{figure*}
\includegraphics[width=\columnwidth]{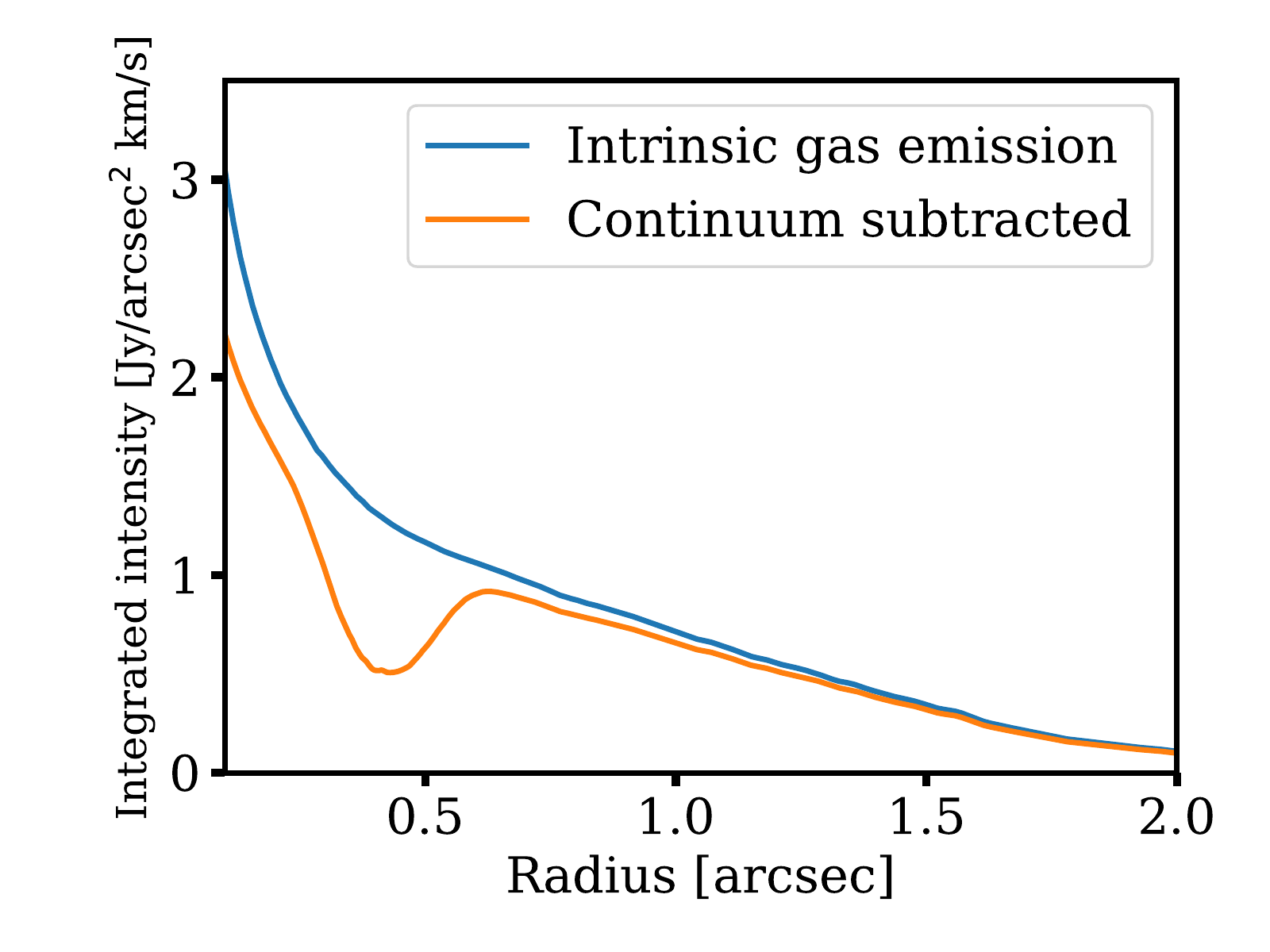}
\includegraphics[width=\columnwidth]{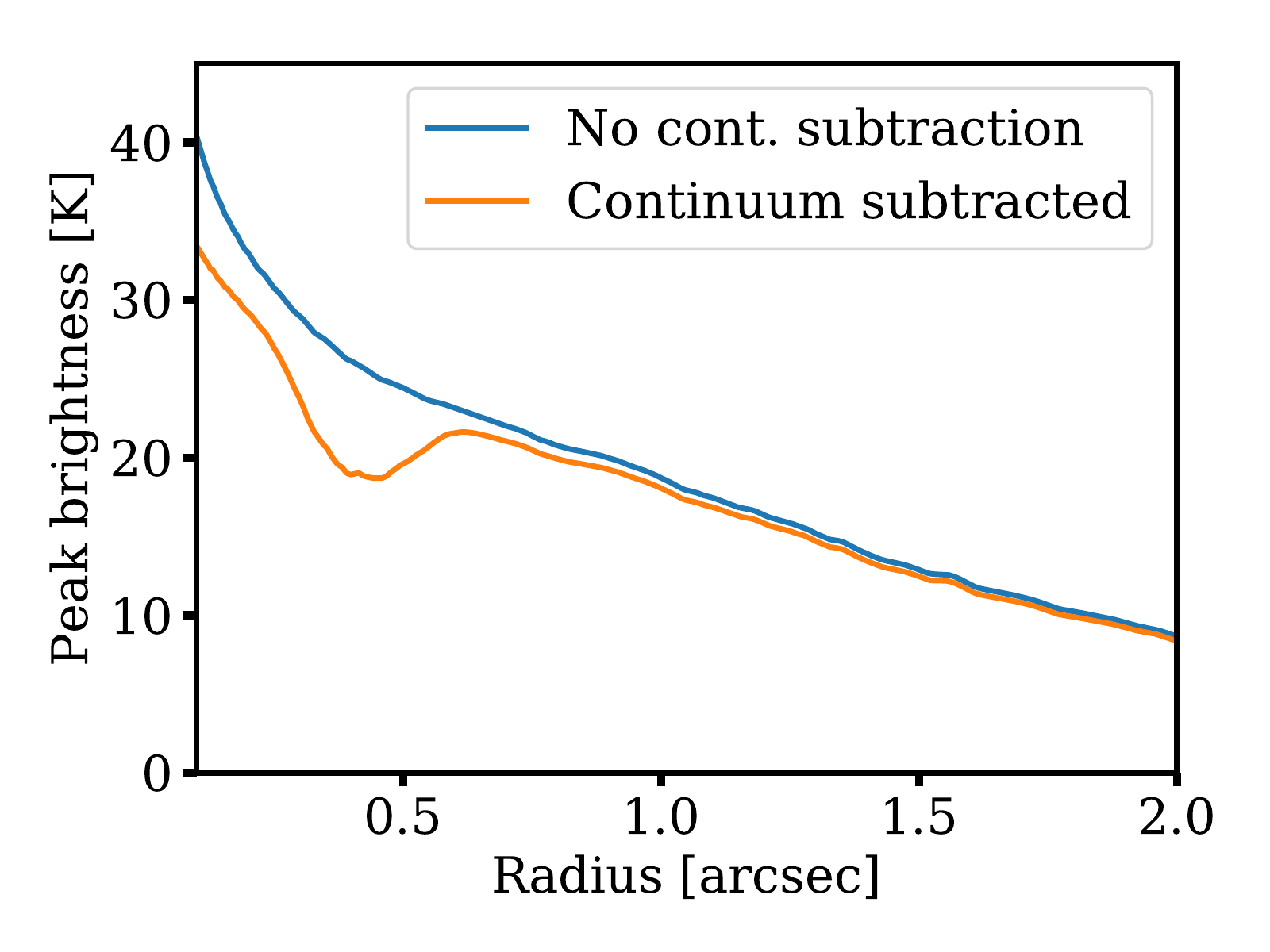}
\caption{Radial profiles for an illustrative radiative transfer model, with a smooth gas surface density profile but with a bright ring in the dust. \textit{Left}: radial profile of the integrated intensity map after performing continuum subtraction. Because the integrated intensity is not defined when not performing continuum subtraction, we compare it with the intrinsic gas emission (i.e., setting the dust opacity to zero), although we note that this is not a quantity that is possible to reconstruct in observations. \textit{Right}: peak brightness map, both for continuum subtracted and non-continuum subtracted case. In contrast to the integrated intensity, both these quantities can be easily be computed from observations. In both cases, the continuum subtracted maps show a spurious gap at the location of the continuum ring, while the non continuum subtracted peak brightness map correctly recovers a smooth gas emission profile.}
\label{fig:spurious_substructure}
\end{figure*}

\subsubsection{Spurious substructure}

We use a disc radiative transfer model to see the difference brought by continuum subtraction in identifying sub-structures. We have prescribed a smooth gas density $\Sigma_\mathrm{gas}$, which follows a power-law with a slope of 0.5: $\Sigma_\mathrm{gas} \propto r^{-0.5}$. For the dust surface density $\Sigma_\mathrm{dust}$ instead, to this power-law profile we have superimposed a Gaussian ring located at $r_0=60$ au with a width $\sigma=10$ au and an amplitude $A=10$ with respect to the background profile: 
\begin{equation}
    \Sigma_\mathrm{dust} (r) \propto \left\{ 1+A \exp\left[-\frac{(r-r_0)^2}{2 \sigma^2}\right]\right\}r^{-0.5}
\end{equation}.

We show in \autoref{fig:spurious_substructure} the emission maps (integrated intensity in the left panel and peak brightness in the right panel). As it can be seen the continuum subtracted integrated intensity and peak brightness maps show an artificial gap if subtracting the continuum (orange lines). To show that this structure is spurious, we have also ray-traced the model removing the contribution to the opacity of the dust and plotted its integrated intensity as the blue line in the left panel. This extra step is necessary because an integrated intensity map is not well defined if continuum subtraction is not performed, since the value would depend on the integration boundaries. In the observational case, therefore, it is not possible to recover the intrinsic gas emission without extensive modelling. For the peak brightness map, instead, the blue line represents the map computed without performing continuum subtraction. This shows that this procedure correctly recovers the fact that there is no sub-structure in the gas emission. The procedure can easily be applied to observational datasets; moreover, as we discuss in appendix \ref{append:continuum_subtraction}, integrated intensity maps are complex because at any given location the gas emission may be optically thick at the line center while being optically thin in the line wings. Peak brightness map are instead easier to interpret because they only capture the highest optical depth part of the emission. For this reason, we will not consider further integrated intensity maps and we will restrict our analysis to peak brightness maps in what follows.

\begin{figure}
\includegraphics[width=\columnwidth]{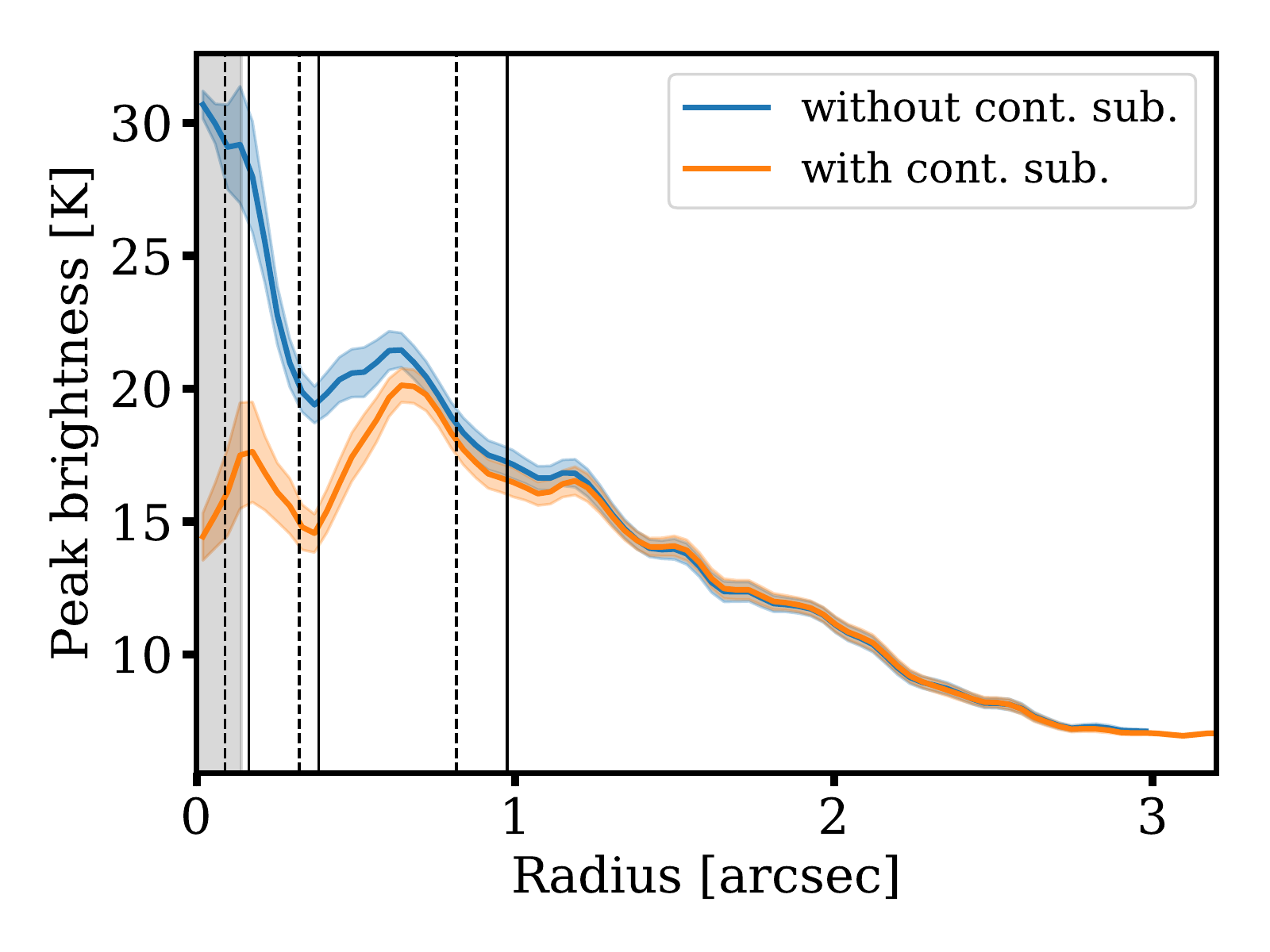}
\caption{Radial profile of the peak intensity map from the $^{13}$CO observations, with (orange) and without (blue) continuum subtraction. If continuum subtraction is not performed, the gap at 0.4 arcsec becomes significantly shallower. However, the gap remains, implying it corresponds to a real feature in the intrinsic gas emission profile. Black vertical lines indicate the location of the continuum features, solid for rings (bright features) and dashed for gaps (dark features). The grey shaded area marks the size of the beam.}
\label{fig:radial_profiles_notsubtracted}
\end{figure}

\subsubsection{Non continuum-subtracted $^{13}$CO profile} 

In light of the discussion above, we show in \autoref{fig:radial_profiles_notsubtracted} the difference between these two approaches for the $^{13}$CO emission on the peak brightness map. As it can be seen, the difference is significant. The gap in $^{13}$CO emission in this way is significantly shallower: the brightness temperature at the local minimum has increased from 15 to 19 K. For reference, the brightness temperature at the local maximum outside the gap is $\sim$21 K. This implies that most of the observed gap in \autoref{fig:radial_profiles} was an artefact of the continuum subtraction. However, we also note that a gap is still visible even in the non-continuum subtracted datacube, implying it is a real dip in emission. Some concern could be raised that the dip may be due to the absorption of the back side of the CO-emitting layer from high optical depth continuum. This can happen only when the (spatial and/or spectral) resolution is poor since at a given spatial location the front and back side fall in two different velocity channels, and the peak brightness map ensures we are selecting the front side emission. Nevertheless, if continuum absorption was responsible for the gap, we should see it in models that correctly reproduce the continuum emission in the rings, such as those we present in section \ref{sec:gap_13co}. Because this is not the case, we discount this explanation. To confirm it even more, we also experimented with radiative transfer models where we artificially increased the continuum absorption, and also in this case we do not see any gap in the models. In section \ref{sec:gap_13co} we will analyse the possible origin of this feature using radiative transfer models.

\subsection{The extent of the disc}
\label{sec:radial_extent}

\begin{figure*}
\includegraphics[width=\columnwidth]{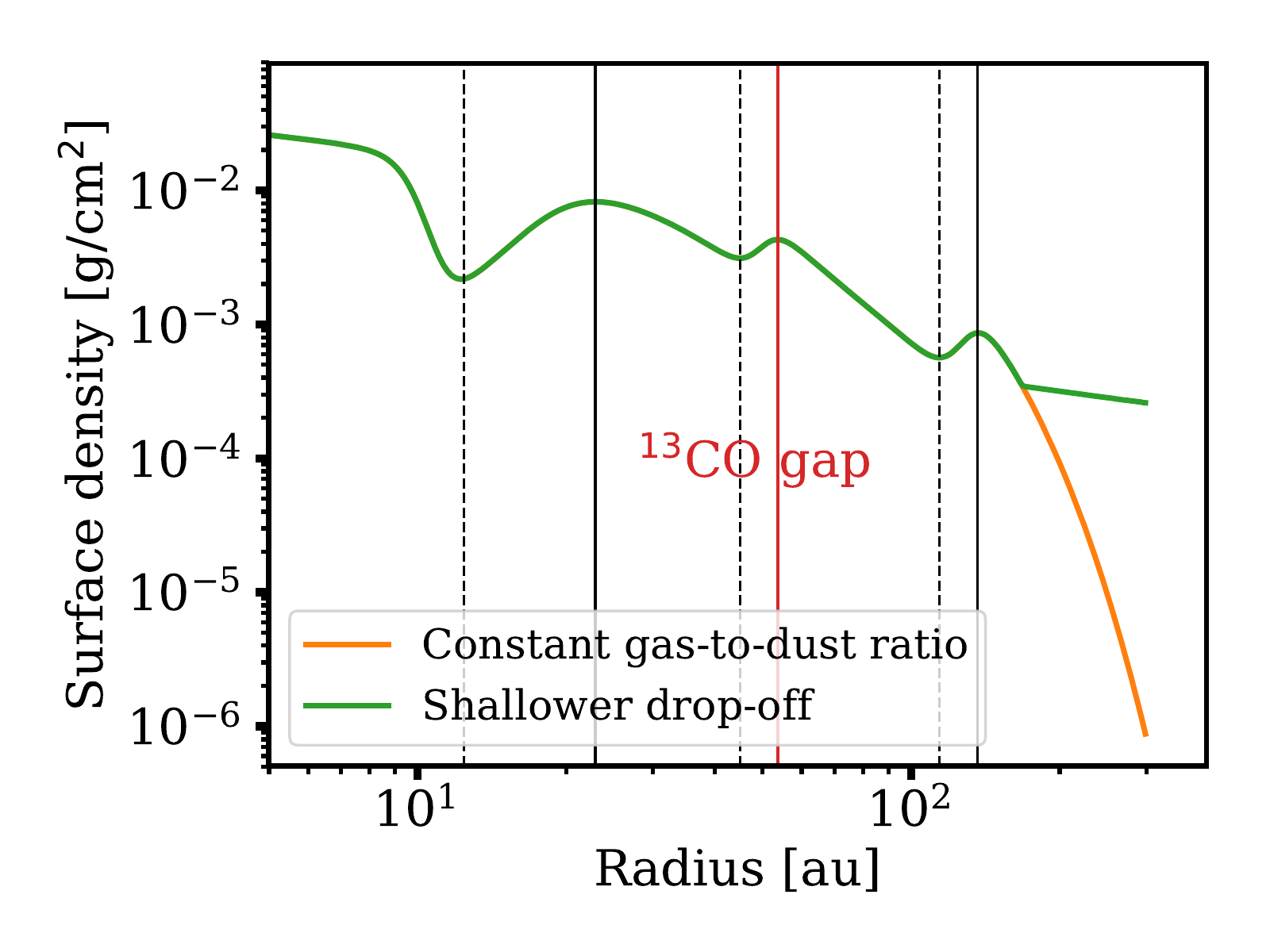}
\includegraphics[width=\columnwidth]{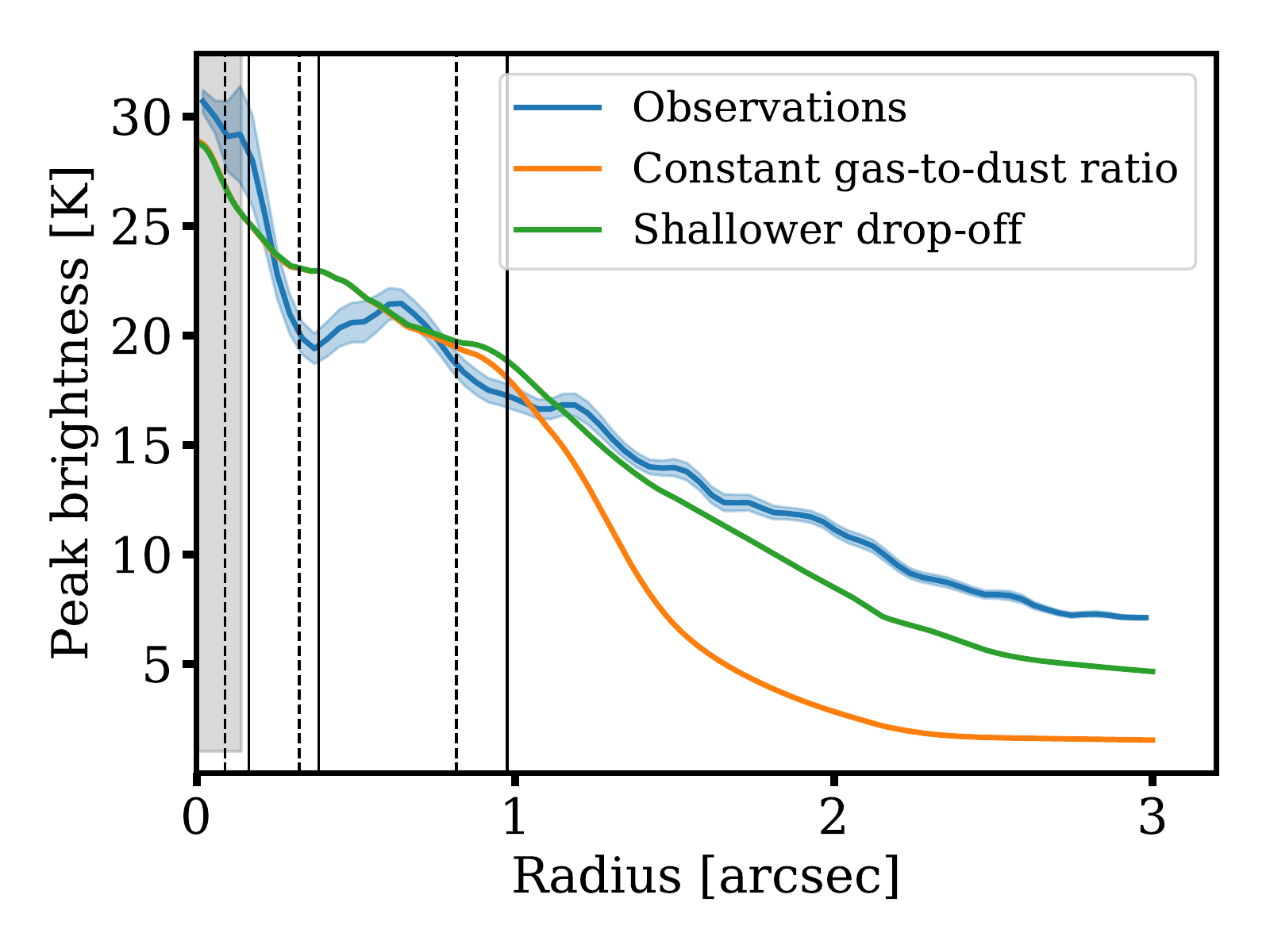}
\caption{\textit{Left:} dust surface density of the ``constant dust-to-gas ratio'' model and of its variant with a shallower drop-off. As in previous plot we have marked the locations of the continuum features, as well as the position of the $^{13}$CO gap. \textit{Right:} predicted $^{13}$CO emission profile of the two models in comparison with the observations. The grey shaded area marks the size of the beam. The vanilla ``constant dust-to-gas ratio'' model clearly exhibits in the outer part of the disc a drop-off of the $^{13}$CO emission that is too steep, implying that the decline of the gas surface density must be shallower than the one of the dust (see text for clarifications on what dust surface density means in this context).}
\label{fig:disc_extent}
\end{figure*}

As mentioned in section \ref{fig:13CO_observations}, the extent of the emission in $^{13}$CO is larger than in the continuum. This is a common feature in proto-planetary discs \citep[e.g.,][]{Isella2007,Panic2009,Isella2012,Ansdell2018} and there is a large body of literature addressing the question of whether this is an opacity or surface density effect \citep[e.g.,][]{Dutrey1998,Hughes2008,Andrews2012,Cleeves2016,Facchini2017,Trapman2019}. In the former interpretation, the difference in size is just apparent and it is due to the gas being optically thick; therefore the emission can be traced at larger radii because the surface density needs to decrease enough for emission to become optically thin. In the latter interpretation, the difference in the observed size reflects a real difference in the gas and dust surface density.

The reason why settling this issue is difficult is because it requires knowing how sharply the continuum profile decreases in the outer part of the disc. Even in the ALMA era, there are only few discs that have been resolved at high spatial resolution (<100 milliarsec) in the continuum, ensuring that the drop-off is spatially resolved, and that also have high-resolution line data. CI Tau has both continuum and $^{13}$CO high-resolution data available; it is therefore instructive to use our high-resolution continuum observations (50 milliarcsec) to set the dust surface density in the radiative transfer model, investigating whether the difference in disc size could be due to the gas opacity. To this end we have taken the best fit to the continuum surface brightness profile presented in \citet{Clarke18} and used it to set both the gas surface density and the surface density of large grains (see section \ref{sec:radiative_transfer}); in what follows we will call this model ``constant gas-to-dust ratio'' (see \autoref{tab:rt_sigma}). We normalise the dust mass to recover the continuum flux at 0.89mm, and assume a dust-to-gas ratio of 100. The dust surface density of the model is shown in the left panel of \autoref{fig:disc_extent} as the orange line.

We show a comparison between this model and the $^{13}$CO observations in the right panel of \autoref{fig:disc_extent}. Within 1 arcsec, the model is a resonable match to the observations; we also show in \autoref{fig:emission_surface} that the emission surface of the model is at a height $z/r \sim 0.1$, in line with what is inferred from the projected velocity map. However, it can be seen how in this model the gas emission drops off too quickly in comparison with the observations beyond $\sim$1 arcsec. For this reason, we run an alternative model where we apply a shallower drop-off of the gas surface density, extending the disc beyond 170 au with a power-law with a slope of 0.5. This procedure is in no way unique and we apply it only for illustrative purposes. The left panel shows (blue line) the surface density we use in the shallower drop-off model and the right panel the predicted emission, showing that the model better reproduces the emission in the outer part of the disc, although a more satisfying fit would require an even shallower drop-off. It is also possible that the external radiation field is slightly warmer than the 10 K we assume in the radiative transfer calculation, slightly increasing the temperature and therefore the CO emission in the outer parts of the disc.

On the basis of these models and of the sharp drop-off resolved in the continuum, we can exclude in this disc the possibility that the difference in size is due only to the gas opacity. This points to a real difference in the dust-to-gas ratio. Note however that in this context ``dust'' is to be intended as the solid component with a significant sub-mm opacity, and not as the \textit{total} solid content. A significant amount of dust locked up in small grains (with size smaller than 100-200 $\mu$m) could be present in the outer disc, but it would remain undetectable in our continuum observations. This is due to what \citet{Rosotti2019a} called ``opacity cliff'', namely the sudden drop in the opacity at millimetre wavelengths for grains smaller than 100-200 $\mu$m. A reduction in grain size is also supported by the variation in the spectral index of the continuum emission (see section \ref{sec:spectral_index}). In the interpretation of \citet{Rosotti2019a}, the sharp drop-off of the dust emission is due to the processes controlling grain growth and therefore the dust opacity, but a significant dust population is still present beyond the continuum outer edge. An alternative interpretation \citep{Birnstiel2014} instead is that radial drift causes a sharp drop-off in the dust-to-gas ratio where the gas profile has a steep pressure gradient. Although our observations do not suggest a steep decline in the gas surface density at the location of the continuum outer edge, addressing this question would require more extensive modelling.  Therefore, both possibilities remain open. 

\subsection{The gap in $^{13}$CO emission}
\label{sec:gap_13co}

\begin{figure*}
\includegraphics[width=\columnwidth]{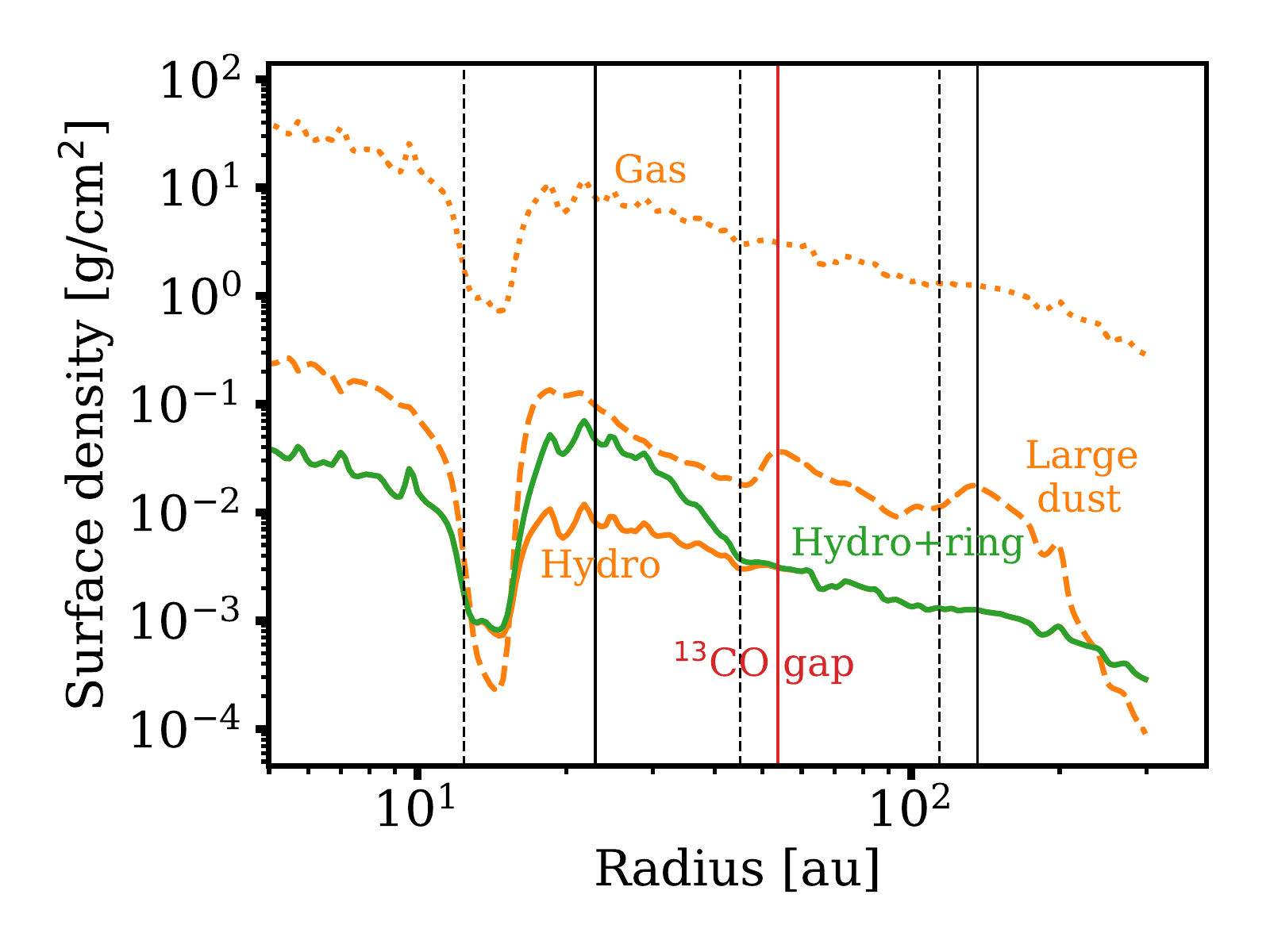}
\includegraphics[width=\columnwidth]{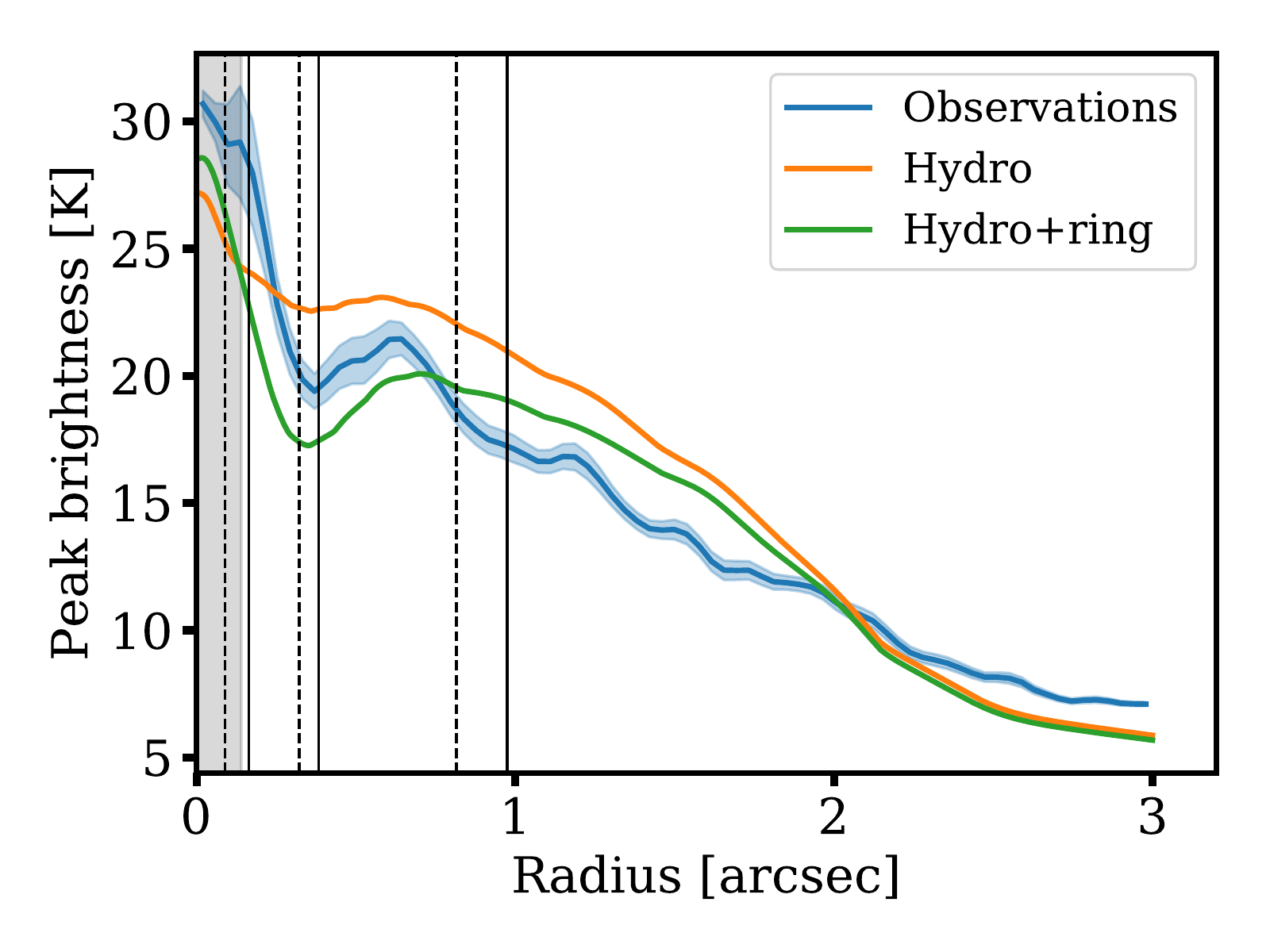}
\caption{\textit{Left:} surface density of gas (dotted line) and dust (both ``small'', solid lines, and ``large'', dashed line, grains) in the ``hydro'' and ``hydro+ring'' models. The ``hydro+ring'' model differs only in the small dust from the ``hydro'' model. The gas surface density has a deep gap at $\sim$15 au and it is essentially smooth further out. To ease the comparison we have marked the positions of the continuum features and of the $^{13}$CO gap. \textit{Right:} predicted $^{13}$CO radial emission profile. The grey shaded area marks the size of the beam.}
\label{fig:comparison_hydro}
\end{figure*}

As mentioned in section \ref{sec:13co}, the $^{13}$CO emission shows a gap in emission at a radius of 0.4 arcseconds. However, the spatial location of this gap does not coincide with the location of a dust gap, but with the location of a dust ring, making it unlikely that this structure is due to a reduction in the gas surface density. Setting aside this issue for a moment, we construct radiative transfer models to establish whether a reduction in surface density comparable to that observed in the continuum would cause a significant reduction in gas emission. To this end, we consider two families of models (see \autoref{tab:rt_sigma}). The first family is the "constant gas-to-dust ratio" family already introduced in the previous section, where we assign a gas surface density rescaling the observed continuum surface density. The second family is the "hydro" family where we use the surface density of the hydrodynamical model containing three planets presented in \citet{Clarke18}. This model also reproduces the continuum observations by construction. In this model, as it is common in dusty planet-disc interaction simulations \citep{Paardekooper2004,Dong2015,Rosotti2016,Dipierro2017}, features in the gas surface density are shallower than in the dust. The only modification we do to the output of the hydro simulation is that in the hydro simulation the surface density profile steepens from $\Sigma_\mathrm{gas} \propto 1/r$ to $1/r^2$ beyond 60 au, but in early tests this produced too little emission beyond this radius. We therefore multiply the gas surface density profile by a factor $r$.

We show the results of this exercise in \autoref{fig:comparison_hydro}. The figure shows the input surface densities (gas and dust) in the left panel and the predicted emission in the right panel for the ``hydro'' model, while the ``constant gas-to-dust ratio'' model has already been shown in \autoref{fig:disc_extent}. Neither of the two models produce a significant gap at the observed location, but they both reproduce correctly the overall surface brightness. This implies that, given the gap depth measured in the dust, the gap depth observed in the gas cannot be caused by a reduction in surface density. This can readily be understood as at the gap location we find that the $^{13}$CO emission is optically thick, with an optical depth at the center of the line of roughly 20. Given that the emission is optically thick, our model is not unique - the optical depth could also be higher, or moderately lower (as long as it is above 1), and still reproduce correctly the observed intensity. However, the fact that we do reproduce the observed intensity confirms that the emission is optically thick. More empirically, this can also be understood as the peak brightness is above 20K, implying that the emission is coming from a layer above that where CO freezes out\footnote{In the models, we find that the temperature at the emission surface is actually higher, roughly 30K at 0.4 arcsec. Comparing the line cubes before and after convolution reveals this is a consequence of beam dilution significantly decreasing the peak brightness in the channel maps. This further reinforces the conclusion that the $^{13}$CO emission is optically thick at these radii.}. This explains why a reduction of a factor $\sim$2 in surface density, as observed in the continuum, is not observable in $^{13}$CO.

We therefore have to seek a different origin for the observed gap, which could explain the gap location. If the gap is not due to a reduction in surface density, it must be due either to a chemical effect (i.e., a change in abundance) or to a reduction in temperature. In what follows we attempt to put forward a possible explanation based on the latter option, although we note that it is certainly not the only possible one.

\begin{figure}
\includegraphics[width=\columnwidth]{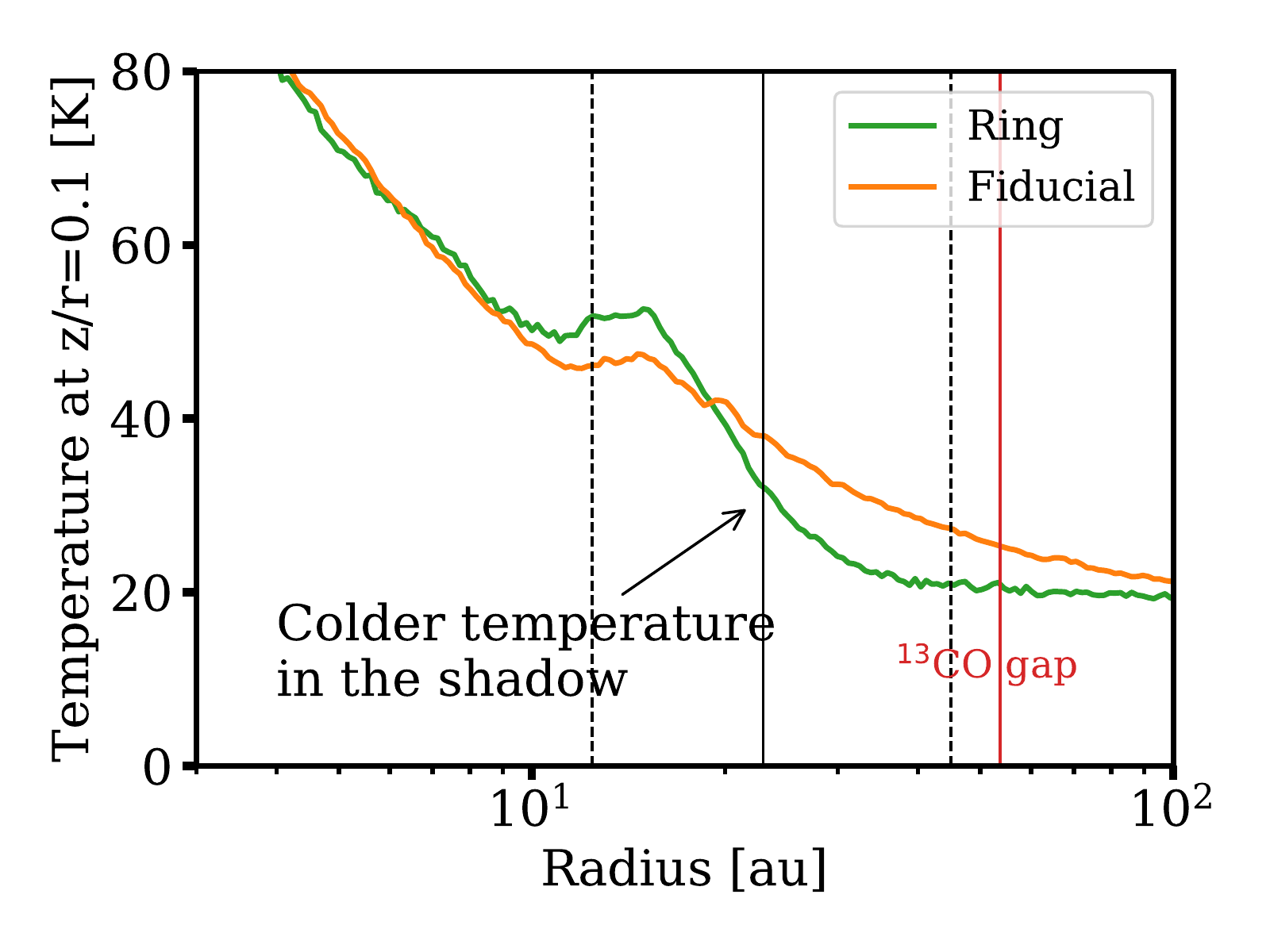}
\caption{Temperature at the emission surface ($z/r=0.1$) for the ``hydro'' and ``hydro+ring'' models. In the ``hydro+ring'' model, at small radii ($\sim$15 au) the temperature increases as the ring intercepts more stellar radiation. At larger radii (20-30 au), however, the ring shadows the outer part of the disc, causing a reduction in disc temperature. This reduction is the reason why the $^{13}$CO emission from this part of the disc is dimmer than in the ``hydro'' model. As in previous plots we have marked the positions of the continuum features and of the $^{13}$CO gap.}
\label{fig:comparison_temperature}
\end{figure}

We hypothesise that the observed gap could be due to shadowing of the disc from dust accumulation at smaller radii. To make this hypothesis more concrete, we have artificially increased the surface density of ``small'' grains in the ``hydro'' model at the location of the first dust ring; we call ``hydro+ring'' the resulting model. This has been accomplished by multiplying the small dust surface density by a Gaussian function $A\exp(-(r-r_0)^2/2\sigma^2)$; we use $r_0=20$ au, $A=5$ and $\sigma=10$ au. Because the ring is only in the ``small'' grains, there is no practically no difference in the continuum emission at sub-mm wavelengths. The result of this exercise on the $^{13}$CO emission is shown in \autoref{fig:comparison_hydro} as the green line. While we do not attempt a detailed fit to the observations, the figure shows that this mechanism can indeed create a gap in the emission, with roughly the correct width and depth. In this scenario the coincidence of the gap with the second dust ring is fortuitous, since in principle the shadow could be cast elsewhere; but we note that our model did not require vast amounts of fine-tuning to correctly reproduce the gap location, and the first ring provides a natural place where to accumulate dust.

As mentioned, in this case the gap is due to a reduction in temperature. This is confirmed by inspecting \autoref{fig:comparison_temperature}, which shows the temperature in the ``hydro'' and ``hydro+ring'' model at the emission surface. At the location of the dust ring, the temperature increases as the ring is exposed to the stellar radiation. Because it is hotter, the ring puffs up and shadows the outer part of the disc, which in turn becomes colder. The situation closely resembles that found at the disc inner rim \citep[e.g.,][]{Dullemond2001}.

We stress that this model should be taken only as illustrative. There is considerable degeneracy regarding whether the ring should be located and what its properties should be; moreover, while here we have used ``small'' (ISM-like) grains, in principle other grain sizes would have a similar effect as long as they create a similar change in the dust optical depth. While clearly degenerate, the model confirms that a temperature reduction is a possible explanation for the observed gap, and that this reduction could be caused by shadowing from the inner disc in a manner similar to our proposed hypothesis.

\section{Discussion}
\label{sec:discussion}

\subsection{A note of caution in analysing high-resolution gas emission maps}

Our radiative transfer model shows that a depletion of the total gas surface density at the location of the gap observed in the continuum is unlikely be responsible for the dip we observe in the $^{13}$CO emission. We proposed that this could be due to shadowing from the inner disc (for example at the location of the first continuum ring), decreasing the disc temperature at larger radii. This is surely not the only possible explanation, with alternatives invoking (for example) chemical depletion of CO by conversion into other species at the location of the continuum ring. While exploring these possibilities is outside the scope of this paper, our results do nevertheless raise a note of caution in analysing high-resolution line emission maps. On one hand, CO is a very convenient tool to use to study the total gas amount due to its high abundance. Its many isotopologues are also conveniently spaced in abundance, so that simultaneous analysis of multiple isotopologues can yield robust upper/lower limits on the total gas column. For these two reasons CO has long been used as a gas mass tracer. On the other hand, our observations show that when analysing spatially resolved data, one should be careful as other effects, such as the issue of continuum subtraction, as well as the temperature at the emitting region, might come into play. We note that in terms of the total emission the effect in our radiative transfer model is limited as it amounts to a difference of only a few Kelvin -- studies interested in measuring the gas mass will therefore not be significantly affected by the issue of continuum subtraction nor by temperature variations. The note of caution we raise is relevant for analysis of sub-structure in discs, i.e. what are colloquially called "gaps" and "rings" -- in this case, mapping the observed surface brightness to a gas surface density is not trivial.

In the future, it will be interesting to compare studies such as this that rely on the surface brightness of line emission with studies that use instead the disc kinematics \citep{Teague2018,Casassus2019}. Disc kinematics can also be used to derive information about the disc substructure, such as the presence and location of pressure maxima \citep{Teague2018,Keppler2019} and the width of gas features \citep{Rosotti2020b}. The two methods are complementary and comparing them will be essential to study disc sub-structures and their origin.

\subsection{On the planets hosted in the disc}

\begin{figure}
\includegraphics[width=\columnwidth]{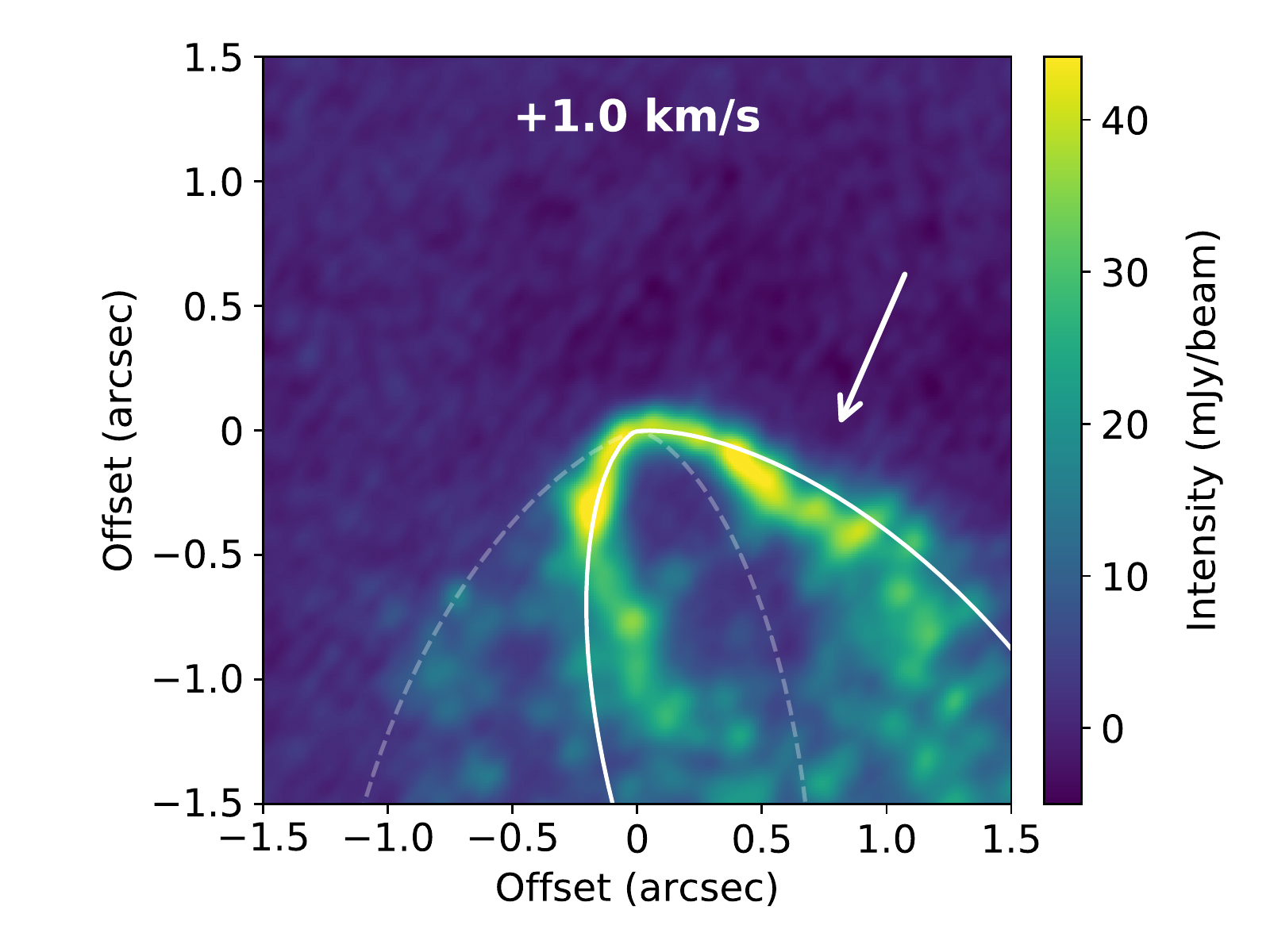}
\caption{A possible deviation from Keplerian rotation seen in the $^{12}$CO channel map at 1 km/s (marked by the arrow) at a deprojected radial distance of $1\farcs3$ (180\,au at 140\,pc).  The solid white line marks the iso-velocity contour for an unperturbed emission surface at $z/r=0.3$, with the corresponding contour for the back side of the disc is shown with the dashed white line.}
\label{fig:12co_kink}
\end{figure}

\citet{Clarke18} proposed that the three gaps observed in the continuum are due to a family of three, Jupiter-mass planets and presented hydro-dynamical models to support this interpretation. Overall, the data we present in this paper neither excludes nor confirms this hypothesis. As our radiative transfer models shows (see \autoref{fig:comparison_hydro}), none of the gaps opened by the planets in the gas surface density would be observable. While for the second and third gap we find that this is because the $^{13}$CO line emission is always optically thick in the gap, for the first, deep gap we find instead that the limiting factor is spatial resolution. Future observations at higher resolution might better probe the planet hypothesis for the first gap; alternatively, observations of rarer, more optically thin species might probe this for the other two gaps.

While we are unable to study the planetary hypothesis for the continuum gaps further, there is a possibility that this system hosts an additional planet. We show this in \autoref{fig:12co_kink}, which shows the $^{12}$CO channel map at 1 km/s. The arrow on the figure marks the location of a possible deviation from Keplerian rotation, which is seen at a deprojected radial distance of $1\farcs3$ (180\,au at 140\,pc). These features have been shown to be compatible with the presence of a massive planet (\citealt{Pinte2018planet,Pinte2019}; see also \citealt{Pinte2020} for other tentative deviations), in which case they correspond to local deviations of the Keplerian rotation pattern induced by the planet gravity. In some cases \citep{Pinte2019,Pinte2020}, the deviations have been found inside a continuum gap, whereas in this case the feature is found outside the extent of the continuum disc, as it is for HD163296 \citep{Pinte2018planet}. At the signal-to-noise of these observations, the detection of this deviation from Keplerian rotation is not unambiguous. Moreover, due to foreground absorption, we cannot check whether the feature appears also in the corresponding channel on the other side of the disc. Future observations are required to confirm the deviation, its properties, and study its origin.

\subsection{The ring of CS emission}

The CS emission shows a single ring-like morphology with no central component.  The emission peaks at a similar location to the second continuum gap and ring at $\sim$50 au.  

A ring morphology is in agreement with observations and modelling of CS in other discs (e.g., \citealt{LeGal2019}), indicating that the ring is likely produced by chemical effects rather than a variation of the gas surface density or gas temperature. Our analysis shows that the ring is slightly offset from the central star, consistent with the CS emission originating from a relative height of $z/r \sim 0.1$ in the disc. \citet{guilloteau_2012} perform chemical modelling to explain observations of CS toward DM Tau.  They find the emission originates from $z/r \sim 0.15$--0.2 (above one scale height), broadly in agreement with their model predictions of a CS reservoir between $z/r \sim 0.2$--0.4.  More recently, \citet{LeGal2019} analysed CS emission from a broader sample of discs (including DM Tau).  While relative heights were not directly measured, their models show agreement with the above, placing CS reservoirs at $z/r > 0.1$.  The relative height of $z/r$ in CI Tau is toward the lower end of these estimates, and may hint that CS is being more efficiently destroyed in the upper layers of the CI Tau disc (e.g. via photodissociation) when compared with other objects. Another possibility involves instead the formation route. While performing chemical modelling is outside the scope of this paper, formation routes of CS have been explored by \citet{Semenov2018} and \citet{LeGal2019}. They both find that CS has two main formation pathways. The first one involves reactions between S$^+$ and small hydro-carbons produced by C$^+$ gas-phase chemistry. The second, slower route involves neutral-neutral reactions of S with molecules such as CH, CH$_2$ and C$_2$. The former route requires high radiation fields to photoionize S and C and can therefore only take place relatively high above the midplane. Instead, the latter route only involves neutral species, and can therefore produce a reservoir of CS at a lower relative height in the disc.  The latter formation route would also be compatible with the scenario presented in section \ref{sec:gap_13co} where this region of the disc is partially shadowed, which would presumably inhibit the first formation route.

\section{Conclusions}
\label{sec:conclusions}

In this paper we presented ALMA band 7 high-resolution (0.15 arcsec) observations of the disc around CI Tau in 0.89mm continuum, $^{12}$CO \& $^{13}$CO $J=3$--2, and CS $J=7$-6 CS emission lines. Our results are as follows:
\begin{itemize}
    \item The morphology of the continuum emission closely resembles our previous observations in band 6 at 1.3mm. The spectral index rises towards the outer parts of the disc, indicating that the disc is slightly smaller at 1.3mm than at 0.89mm. This might be linked to a sharp variation in grain size at the location of the continuum outer edge.
    
    \item The $^{12}$CO is heavily contaminated by foreground absorption, preventing any analysis of the disc structure using this emission line. 
        
    \item The $^{13}$CO emission shows a gap in emission coincident with the location of the second continuum ring (rather than a gap) at a radius of $\sim$ 50 au. Most of this gap is likely an artefact resulting from continuum subtraction, since the gas emission is optically thick. However, the gap still remains even if continuum subtraction is not performed, suggesting that at least partially the gap in emission is real (though less conspicuous than suggested by the continuum-subtracted map).
    
    \item We propose that the inner disc (radius $\lesssim$ 20 au) is shadowing the outer part of the disc, decreasing the temperature and therefore creating at a radius of $\sim$ 50 au the observed gap in emission, that does not however correspond to a decrease in surface density. This raises a note of caution in mapping high-resolution observations of gas emission lines to the underlying gas surface density.
    
    \item The $^{13}$CO emission is more extended than the continuum disc. Given the sharp drop-off observed in the continuum emission, this is difficult to reconcile with an opacity effect and is more likely due to a genuine difference in the dust and gas density distributions.
    
    \item The CS emission is resolved in a ring peaking at $\sim$50 au. Based on previous observational and modelling studies, this is most likely due to chemical effects rather than a variation in gas surface density.

\end{itemize}

\section*{Acknowledgements}

We thank an anonymous referee for a careful reading of the manuscript and many suggestions that improved the quality of this work. We would like to thank Richard Teague for helpful discussions regarding the \texttt{GoFish} package and Ryan Loomis for advice on the residual scaling. This work has been supported by the DISCSIM project, grant agreement 341137 funded by the European Research Council under ERC-2013-ADG. GR acknowledges support from the Netherlands Organisation for Scientific Research (NWO, program number 016.Veni.192.233). JDI acknowledges support from the STFC under ST/R000549/1 and ST/T000287/1. SF acknowledges an ESO Fellowship. MT, RB and CC acknowledge support from the STFC consolidated grant ST/S000623/1. This work has also been supported by the European Union’s Horizon 2020 research and innovation programme under the Marie Sklodowska-Curie grant agreement No 823823 (DUSTBUSTERS). MK gratefully acknowledges funding by the University of Tartu ASTRA project 2014-2020.4.01.16-0029 KOMEET, financed by the EU European Regional Development Fund. This paper makes use of the following ALMA data: ADS/JAO.ALMA\#2017.A.00014.S, ADS/JAO.ALMA\#2016.1.01370.S. ALMA is a partnership of ESO (representing its member states), NSF (USA) and NINS (Japan), together with NRC (Canada), MOST and ASIAA (Taiwan), and KASI (Republic of Korea), in cooperation with the Republic of Chile. The Joint ALMA Observatory is operated by ESO, AUI/NRAO and NAOJ.

\section*{Data availability}
The data underlying this paper are available in the ALMA archive: ADS/JAO.ALMA\#2017.A.00014.S.




\bibliographystyle{mnras}
\bibliography{ci_tau_7}

\begin{thebibliography}{}
\makeatletter
\relax
\def\mn@urlcharsother{\let\do\@makeother \do\$\do\&\do\#\do\^\do\_\do\%\do\~}
\def\mn@doi{\begingroup\mn@urlcharsother \@ifnextchar [ {\mn@doi@}
  {\mn@doi@[]}}
\def\mn@doi@[#1]#2{\def\@tempa{#1}\ifx\@tempa\@empty \href
  {http://dx.doi.org/#2} {doi:#2}\else \href {http://dx.doi.org/#2} {#1}\fi
  \endgroup}
\def\mn@eprint#1#2{\mn@eprint@#1:#2::\@nil}
\def\mn@eprint@arXiv#1{\href {http://arxiv.org/abs/#1} {{\tt arXiv:#1}}}
\def\mn@eprint@dblp#1{\href {http://dblp.uni-trier.de/rec/bibtex/#1.xml}
  {dblp:#1}}
\def\mn@eprint@#1:#2:#3:#4\@nil{\def\@tempa {#1}\def\@tempb {#2}\def\@tempc
  {#3}\ifx \@tempc \@empty \let \@tempc \@tempb \let \@tempb \@tempa \fi \ifx
  \@tempb \@empty \def\@tempb {arXiv}\fi \@ifundefined
  {mn@eprint@\@tempb}{\@tempb:\@tempc}{\expandafter \expandafter \csname
  mn@eprint@\@tempb\endcsname \expandafter{\@tempc}}}

\bibitem[\protect\citeauthoryear{{ALMA Partnership} et~al.,}{{ALMA Partnership}
  et~al.}{2015}]{HLTau}
{ALMA Partnership} et~al., 2015, \mn@doi [\apjl] {10.1088/2041-8205/808/1/L3},
  \href {https://ui.adsabs.harvard.edu/abs/2015ApJ...808L...3A} {808, L3}

\bibitem[\protect\citeauthoryear{{Andrews} \& {Williams}}{{Andrews} \&
  {Williams}}{2005}]{2005ApJ...631.1134A}
{Andrews} S.~M.,  {Williams} J.~P.,  2005, \mn@doi [\apj] {10.1086/432712},
  \href {https://ui.adsabs.harvard.edu/abs/2005ApJ...631.1134A} {631, 1134}

\bibitem[\protect\citeauthoryear{{Andrews} et~al.,}{{Andrews}
  et~al.}{2012}]{Andrews2012}
{Andrews} S.~M.,  et~al., 2012, \mn@doi [\apj] {10.1088/0004-637X/744/2/162},
  \href {https://ui.adsabs.harvard.edu/abs/2012ApJ...744..162A} {744, 162}

\bibitem[\protect\citeauthoryear{{Andrews} et~al.,}{{Andrews}
  et~al.}{2016}]{Andrews2016}
{Andrews} S.~M.,  et~al., 2016, \mn@doi [\apjl] {10.3847/2041-8205/820/2/L40},
  \href {https://ui.adsabs.harvard.edu/abs/2016ApJ...820L..40A} {820, L40}

\bibitem[\protect\citeauthoryear{{Andrews} et~al.,}{{Andrews}
  et~al.}{2018}]{Andrews2018}
{Andrews} S.~M.,  et~al., 2018, \mn@doi [\apjl] {10.3847/2041-8213/aaf741},
  \href {https://ui.adsabs.harvard.edu/abs/2018ApJ...869L..41A} {869, L41}

\bibitem[\protect\citeauthoryear{{Ansdell} et~al.,}{{Ansdell}
  et~al.}{2018}]{Ansdell2018}
{Ansdell} M.,  et~al., 2018, \mn@doi [\apj] {10.3847/1538-4357/aab890}, \href
  {https://ui.adsabs.harvard.edu/abs/2018ApJ...859...21A} {859, 21}

\bibitem[\protect\citeauthoryear{{Bergner}, {{\"O}berg}, {Bergin}, {Loomis},
  {Pegues}  \& {Qi}}{{Bergner} et~al.}{2019}]{Bergner2019}
{Bergner} J.~B.,  {{\"O}berg} K.~I.,  {Bergin} E.~A.,  {Loomis} R.~A.,
  {Pegues} J.,   {Qi} C.,  2019, \mn@doi [\apj] {10.3847/1538-4357/ab141e},
  \href {https://ui.adsabs.harvard.edu/abs/2019ApJ...876...25B} {876, 25}

\bibitem[\protect\citeauthoryear{{Birnstiel} \& {Andrews}}{{Birnstiel} \&
  {Andrews}}{2014}]{Birnstiel2014}
{Birnstiel} T.,  {Andrews} S.~M.,  2014, \mn@doi [\apj]
  {10.1088/0004-637X/780/2/153}, \href
  {https://ui.adsabs.harvard.edu/abs/2014ApJ...780..153B} {780, 153}

\bibitem[\protect\citeauthoryear{{Boehler}, {Weaver}, {Isella}, {Ricci},
  {Grady}, {Carpenter}  \& {Perez}}{{Boehler} et~al.}{2017}]{Boehler2017}
{Boehler} Y.,  {Weaver} E.,  {Isella} A.,  {Ricci} L.,  {Grady} C.,
  {Carpenter} J.,   {Perez} L.,  2017, \mn@doi [\apj]
  {10.3847/1538-4357/aa696c}, \href
  {https://ui.adsabs.harvard.edu/abs/2017ApJ...840...60B} {840, 60}

\bibitem[\protect\citeauthoryear{{Boehler} et~al.,}{{Boehler}
  et~al.}{2018}]{Boehler2018}
{Boehler} Y.,  et~al., 2018, \mn@doi [\apj] {10.3847/1538-4357/aaa19c}, \href
  {https://ui.adsabs.harvard.edu/abs/2018ApJ...853..162B} {853, 162}

\bibitem[\protect\citeauthoryear{{Carrasco-Gonz{\'a}lez}
  et~al.,}{{Carrasco-Gonz{\'a}lez} et~al.}{2019}]{2019ApJ...883...71C}
{Carrasco-Gonz{\'a}lez} C.,  et~al., 2019, \mn@doi [\apj]
  {10.3847/1538-4357/ab3d33}, \href
  {https://ui.adsabs.harvard.edu/abs/2019ApJ...883...71C} {883, 71}

\bibitem[\protect\citeauthoryear{{Casassus} \& {P{\'e}rez}}{{Casassus} \&
  {P{\'e}rez}}{2019}]{Casassus2019}
{Casassus} S.,  {P{\'e}rez} S.,  2019, \mn@doi [\apjl]
  {10.3847/2041-8213/ab4425}, \href
  {https://ui.adsabs.harvard.edu/abs/2019ApJ...883L..41C} {883, L41}

\bibitem[\protect\citeauthoryear{{Casassus} et~al.,}{{Casassus}
  et~al.}{2013}]{Casassus2013}
{Casassus} S.,  et~al., 2013, \mn@doi [\nat] {10.1038/nature11769}, \href
  {https://ui.adsabs.harvard.edu/abs/2013Natur.493..191C} {493, 191}

\bibitem[\protect\citeauthoryear{{Cazzoletti} et~al.,}{{Cazzoletti}
  et~al.}{2018}]{Cazzoletti2018}
{Cazzoletti} P.,  et~al., 2018, \mn@doi [\aap] {10.1051/0004-6361/201834006},
  \href {https://ui.adsabs.harvard.edu/abs/2018A&A...619A.161C} {619, A161}

\bibitem[\protect\citeauthoryear{{Clarke} et~al.,}{{Clarke}
  et~al.}{2018}]{Clarke18}
{Clarke} C.~J.,  et~al., 2018, \mn@doi [\apj] {10.3847/2041-8213/aae36b}, \href
  {https://ui.adsabs.harvard.edu/\#abs/2018ApJ...866L...6C} {866, L6}

\bibitem[\protect\citeauthoryear{{Cleeves}, {{\"O}berg}, {Wilner}, {Huang},
  {Loomis}, {Andrews}  \& {Czekala}}{{Cleeves} et~al.}{2016}]{Cleeves2016}
{Cleeves} L.~I.,  {{\"O}berg} K.~I.,  {Wilner} D.~J.,  {Huang} J.,  {Loomis}
  R.~A.,  {Andrews} S.~M.,   {Czekala} I.,  2016, \mn@doi [\apj]
  {10.3847/0004-637X/832/2/110}, \href
  {https://ui.adsabs.harvard.edu/abs/2016ApJ...832..110C} {832, 110}

\bibitem[\protect\citeauthoryear{{Dent}, {Pinte}, {Cortes}, {M{\'e}nard},
  {Hales}, {Fomalont}  \& {de Gregorio-Monsalvo}}{{Dent}
  et~al.}{2019}]{Dent2019}
{Dent} W.~R.~F.,  {Pinte} C.,  {Cortes} P.~C.,  {M{\'e}nard} F.,  {Hales} A.,
  {Fomalont} E.,   {de Gregorio-Monsalvo} I.,  2019, \mn@doi [\mnras]
  {10.1093/mnrasl/sly181}, \href
  {https://ui.adsabs.harvard.edu/abs/2019MNRAS.482L..29D} {482, L29}

\bibitem[\protect\citeauthoryear{{Dipierro} \& {Laibe}}{{Dipierro} \&
  {Laibe}}{2017}]{Dipierro2017}
{Dipierro} G.,  {Laibe} G.,  2017, \mn@doi [\mnras] {10.1093/mnras/stx977},
  \href {https://ui.adsabs.harvard.edu/abs/2017MNRAS.469.1932D} {469, 1932}

\bibitem[\protect\citeauthoryear{{Dong}, {Zhu}  \& {Whitney}}{{Dong}
  et~al.}{2015}]{Dong2015}
{Dong} R.,  {Zhu} Z.,   {Whitney} B.,  2015, \mn@doi [\apj]
  {10.1088/0004-637X/809/1/93}, \href
  {https://ui.adsabs.harvard.edu/abs/2015ApJ...809...93D} {809, 93}

\bibitem[\protect\citeauthoryear{{Dullemond}, {Dominik}  \&
  {Natta}}{{Dullemond} et~al.}{2001}]{Dullemond2001}
{Dullemond} C.~P.,  {Dominik} C.,   {Natta} A.,  2001, \mn@doi [\apj]
  {10.1086/323057}, \href
  {https://ui.adsabs.harvard.edu/abs/2001ApJ...560..957D} {560, 957}

\bibitem[\protect\citeauthoryear{{Dullemond} et~al.,}{{Dullemond}
  et~al.}{2018}]{Dullemond2018}
{Dullemond} C.~P.,  et~al., 2018, \mn@doi [\apjl] {10.3847/2041-8213/aaf742},
  \href {https://ui.adsabs.harvard.edu/abs/2018ApJ...869L..46D} {869, L46}

\bibitem[\protect\citeauthoryear{{Dutrey}, {Guilloteau}, {Prato}, {Simon},
  {Duvert}, {Schuster}  \& {Menard}}{{Dutrey} et~al.}{1998}]{Dutrey1998}
{Dutrey} A.,  {Guilloteau} S.,  {Prato} L.,  {Simon} M.,  {Duvert} G.,
  {Schuster} K.,   {Menard} F.,  1998, \aap, \href
  {https://ui.adsabs.harvard.edu/abs/1998A&A...338L..63D} {338, L63}

\bibitem[\protect\citeauthoryear{{Eriksson}, {Johansen}  \& {Liu}}{{Eriksson}
  et~al.}{2020}]{Eriksson2020}
{Eriksson} L. E.~J.,  {Johansen} A.,   {Liu} B.,  2020, \mn@doi [\aap]
  {10.1051/0004-6361/201937037}, \href
  {https://ui.adsabs.harvard.edu/abs/2020A&A...635A.110E} {635, A110}

\bibitem[\protect\citeauthoryear{{Facchini}, {Birnstiel}, {Bruderer}  \& {van
  Dishoeck}}{{Facchini} et~al.}{2017}]{Facchini2017}
{Facchini} S.,  {Birnstiel} T.,  {Bruderer} S.,   {van Dishoeck} E.~F.,  2017,
  \mn@doi [\aap] {10.1051/0004-6361/201630329}, \href
  {https://ui.adsabs.harvard.edu/abs/2017A&A...605A..16F} {605, A16}

\bibitem[\protect\citeauthoryear{{Facchini}, {Pinilla}, {van Dishoeck}  \& {de
  Juan Ovelar}}{{Facchini} et~al.}{2018}]{Facchini2018}
{Facchini} S.,  {Pinilla} P.,  {van Dishoeck} E.~F.,   {de Juan Ovelar} M.,
  2018, \mn@doi [\aap] {10.1051/0004-6361/201731390}, \href
  {https://ui.adsabs.harvard.edu/abs/2018A&A...612A.104F} {612, A104}

\bibitem[\protect\citeauthoryear{{Facchini} et~al.,}{{Facchini}
  et~al.}{2019}]{Facchini2019}
{Facchini} S.,  et~al., 2019, \mn@doi [\aap] {10.1051/0004-6361/201935496},
  \href {https://ui.adsabs.harvard.edu/abs/2019A&A...626L...2F} {626, L2}

\bibitem[\protect\citeauthoryear{{Favre} et~al.,}{{Favre}
  et~al.}{2019}]{Favre2019}
{Favre} C.,  et~al., 2019, \mn@doi [\apj] {10.3847/1538-4357/aaf80c}, \href
  {https://ui.adsabs.harvard.edu/abs/2019ApJ...871..107F} {871, 107}

\bibitem[\protect\citeauthoryear{{Fedele} et~al.,}{{Fedele}
  et~al.}{2017}]{Fedele2017}
{Fedele} D.,  et~al., 2017, \mn@doi [\aap] {10.1051/0004-6361/201629860}, \href
  {https://ui.adsabs.harvard.edu/abs/2017A&A...600A..72F} {600, A72}

\bibitem[\protect\citeauthoryear{{Flock}, {Ruge}, {Dzyurkevich}, {Henning},
  {Klahr}  \& {Wolf}}{{Flock} et~al.}{2015}]{Flock2015}
{Flock} M.,  {Ruge} J.~P.,  {Dzyurkevich} N.,  {Henning} T.,  {Klahr} H.,
  {Wolf} S.,  2015, \mn@doi [\aap] {10.1051/0004-6361/201424693}, \href
  {https://ui.adsabs.harvard.edu/abs/2015A&A...574A..68F} {574, A68}

\bibitem[\protect\citeauthoryear{{Foreman-Mackey}, {Hogg}, {Lang}  \&
  {Goodman}}{{Foreman-Mackey} et~al.}{2013}]{EmceePaper}
{Foreman-Mackey} D.,  {Hogg} D.~W.,  {Lang} D.,   {Goodman} J.,  2013, \mn@doi
  [Publications of the Astronomical Society of the Pacific] {10.1086/670067},
  \href {https://ui.adsabs.harvard.edu/\#abs/2013PASP..125..306F} {125, 306}

\bibitem[\protect\citeauthoryear{{Gaia Collaboration} et~al.,}{{Gaia
  Collaboration} et~al.}{2018}]{gaiadr2}
{Gaia Collaboration} et~al., 2018, \mn@doi [\aap]
  {10.1051/0004-6361/201833051}, \href
  {https://ui.adsabs.harvard.edu/abs/2018A&A...616A...1G} {616, A1}

\bibitem[\protect\citeauthoryear{{Guidi} et~al.,}{{Guidi}
  et~al.}{2016}]{2016A&A...588A.112G}
{Guidi} G.,  et~al., 2016, \mn@doi [\aap] {10.1051/0004-6361/201527516}, \href
  {https://ui.adsabs.harvard.edu/abs/2016A&A...588A.112G} {588, A112}

\bibitem[\protect\citeauthoryear{{Guilloteau}, {Dutrey}, {Pi{\'e}tu}  \&
  {Boehler}}{{Guilloteau} et~al.}{2011}]{Guilloteau2011}
{Guilloteau} S.,  {Dutrey} A.,  {Pi{\'e}tu} V.,   {Boehler} Y.,  2011, \mn@doi
  [\aap] {10.1051/0004-6361/201015209}, \href
  {https://ui.adsabs.harvard.edu/abs/2011A&A...529A.105G} {529, A105}

\bibitem[\protect\citeauthoryear{{Guilloteau}, {Dutrey}, {Wakelam}, {Hersant},
  {Semenov}, {Chapillon}, {Henning}  \& {Pi{\'e}tu}}{{Guilloteau}
  et~al.}{2012}]{guilloteau_2012}
{Guilloteau} S.,  {Dutrey} A.,  {Wakelam} V.,  {Hersant} F.,  {Semenov} D.,
  {Chapillon} E.,  {Henning} T.,   {Pi{\'e}tu} V.,  2012, \mn@doi [\aap]
  {10.1051/0004-6361/201220331}, \href
  {https://ui.adsabs.harvard.edu/abs/2012A&A...548A..70G} {548, A70}

\bibitem[\protect\citeauthoryear{{Guilloteau}, {Simon}, {Pi{\'e}tu}, {Di
  Folco}, {Dutrey}, {Prato}  \& {Chapillon}}{{Guilloteau}
  et~al.}{2014}]{Guilloteau2014}
{Guilloteau} S.,  {Simon} M.,  {Pi{\'e}tu} V.,  {Di Folco} E.,  {Dutrey} A.,
  {Prato} L.,   {Chapillon} E.,  2014, \mn@doi [\aap]
  {10.1051/0004-6361/201423765}, \href
  {https://ui.adsabs.harvard.edu/abs/2014A&A...567A.117G} {567, A117}

\bibitem[\protect\citeauthoryear{{Guilloteau} et~al.,}{{Guilloteau}
  et~al.}{2016}]{Guilloteau2016}
{Guilloteau} S.,  et~al., 2016, \mn@doi [\aap] {10.1051/0004-6361/201527088},
  \href {https://ui.adsabs.harvard.edu/abs/2016A&A...592A.124G} {592, A124}

\bibitem[\protect\citeauthoryear{{Huang} et~al.,}{{Huang}
  et~al.}{2018a}]{Huang2018TwHya}
{Huang} J.,  et~al., 2018a, \mn@doi [\apj] {10.3847/1538-4357/aaa1e7}, \href
  {https://ui.adsabs.harvard.edu/abs/2018ApJ...852..122H} {852, 122}

\bibitem[\protect\citeauthoryear{{Huang} et~al.,}{{Huang}
  et~al.}{2018b}]{Huang2018rings}
{Huang} J.,  et~al., 2018b, \mn@doi [\apjl] {10.3847/2041-8213/aaf740}, \href
  {https://ui.adsabs.harvard.edu/abs/2018ApJ...869L..42H} {869, L42}

\bibitem[\protect\citeauthoryear{{Huang} et~al.,}{{Huang}
  et~al.}{2020}]{2020ApJ...891...48H}
{Huang} J.,  et~al., 2020, \mn@doi [\apj] {10.3847/1538-4357/ab711e}, \href
  {https://ui.adsabs.harvard.edu/abs/2020ApJ...891...48H} {891, 48}

\bibitem[\protect\citeauthoryear{{Hughes}, {Wilner}, {Qi}  \&
  {Hogerheijde}}{{Hughes} et~al.}{2008}]{Hughes2008}
{Hughes} A.~M.,  {Wilner} D.~J.,  {Qi} C.,   {Hogerheijde} M.~R.,  2008,
  \mn@doi [\apj] {10.1086/586730}, \href
  {https://ui.adsabs.harvard.edu/abs/2008ApJ...678.1119H} {678, 1119}

\bibitem[\protect\citeauthoryear{{Isella}, {Testi}, {Natta}, {Neri}, {Wilner}
  \& {Qi}}{{Isella} et~al.}{2007}]{Isella2007}
{Isella} A.,  {Testi} L.,  {Natta} A.,  {Neri} R.,  {Wilner} D.,   {Qi} C.,
  2007, \mn@doi [\aap] {10.1051/0004-6361:20077385}, \href
  {https://ui.adsabs.harvard.edu/abs/2007A&A...469..213I} {469, 213}

\bibitem[\protect\citeauthoryear{{Isella}, {P{\'e}rez}  \&
  {Carpenter}}{{Isella} et~al.}{2012}]{Isella2012}
{Isella} A.,  {P{\'e}rez} L.~M.,   {Carpenter} J.~M.,  2012, \mn@doi [\apj]
  {10.1088/0004-637X/747/2/136}, \href
  {https://ui.adsabs.harvard.edu/abs/2012ApJ...747..136I} {747, 136}

\bibitem[\protect\citeauthoryear{{Isella} et~al.,}{{Isella}
  et~al.}{2016}]{Isella2016}
{Isella} A.,  et~al., 2016, \mn@doi [\prl] {10.1103/PhysRevLett.117.251101},
  \href {https://ui.adsabs.harvard.edu/abs/2016PhRvL.117y1101I} {117, 251101}

\bibitem[\protect\citeauthoryear{{Johansen}, {Oishi}, {Mac Low}, {Klahr},
  {Henning}  \& {Youdin}}{{Johansen} et~al.}{2007}]{Johansen2007}
{Johansen} A.,  {Oishi} J.~S.,  {Mac Low} M.-M.,  {Klahr} H.,  {Henning} T.,
  {Youdin} A.,  2007, \mn@doi [\nat] {10.1038/nature06086}, \href
  {https://ui.adsabs.harvard.edu/abs/2007Natur.448.1022J} {448, 1022}

\bibitem[\protect\citeauthoryear{{Jorsater} \& {van Moorsel}}{{Jorsater} \&
  {van Moorsel}}{1995}]{JvM_1995}
{Jorsater} S.,  {van Moorsel} G.~A.,  1995, \mn@doi [\aj] {10.1086/117668},
  \href {https://ui.adsabs.harvard.edu/abs/1995AJ....110.2037J} {110, 2037}

\bibitem[\protect\citeauthoryear{{Keppler} et~al.,}{{Keppler}
  et~al.}{2019}]{Keppler2019}
{Keppler} M.,  et~al., 2019, \mn@doi [\aap] {10.1051/0004-6361/201935034},
  \href {https://ui.adsabs.harvard.edu/abs/2019A&A...625A.118K} {625, A118}

\bibitem[\protect\citeauthoryear{{Le Gal}, {{\"O}berg}, {Loomis}, {Pegues}  \&
  {Bergner}}{{Le Gal} et~al.}{2019}]{LeGal2019}
{Le Gal} R.,  {{\"O}berg} K.~I.,  {Loomis} R.~A.,  {Pegues} J.,   {Bergner}
  J.~B.,  2019, \mn@doi [\apj] {10.3847/1538-4357/ab1416}, \href
  {https://ui.adsabs.harvard.edu/abs/2019ApJ...876...72L} {876, 72}

\bibitem[\protect\citeauthoryear{{Long} et~al.,}{{Long} et~al.}{2018}]{Long18}
{Long} F.,  et~al., 2018, \mn@doi [\apj] {10.3847/1538-4357/aae8e1}, \href
  {https://ui.adsabs.harvard.edu/\#abs/2018ApJ...869...17L} {869, 17}

\bibitem[\protect\citeauthoryear{{Long} et~al.,}{{Long}
  et~al.}{2019}]{Long2019}
{Long} F.,  et~al., 2019, \mn@doi [\apj] {10.3847/1538-4357/ab2d2d}, \href
  {https://ui.adsabs.harvard.edu/abs/2019ApJ...882...49L} {882, 49}

\bibitem[\protect\citeauthoryear{{Long} et~al.,}{{Long}
  et~al.}{2020}]{Long2020}
{Long} F.,  et~al., 2020, \mn@doi [\apj] {10.3847/1538-4357/ab9a54}, \href
  {https://ui.adsabs.harvard.edu/abs/2020ApJ...898...36L} {898, 36}

\bibitem[\protect\citeauthoryear{{Mac{\'\i}as} et~al.,}{{Mac{\'\i}as}
  et~al.}{2019}]{Macias2019}
{Mac{\'\i}as} E.,  et~al., 2019, \mn@doi [\apj] {10.3847/1538-4357/ab31a2},
  \href {https://ui.adsabs.harvard.edu/abs/2019ApJ...881..159M} {881, 159}

\bibitem[\protect\citeauthoryear{{Morbidelli}}{{Morbidelli}}{2020}]{Morbidelli2020}
{Morbidelli} A.,  2020, \mn@doi [\aap] {10.1051/0004-6361/202037983}, \href
  {https://ui.adsabs.harvard.edu/abs/2020A&A...638A...1M} {638, A1}

\bibitem[\protect\citeauthoryear{{Natta} \& {Testi}}{{Natta} \&
  {Testi}}{2004}]{Natta:2004yu}
{Natta} A.,  {Testi} L.,  2004, in {Johnstone} D.,  {Adams} F.~C.,  {Lin}
  D.~N.~C.,  {Neufeeld} D.~A.,   {Ostriker} E.~C.,  eds,  Astronomical Society
  of the Pacific Conference Series Vol. 323, Star Formation in the Interstellar
  Medium: In Honor of David Hollenbach. p.~279

\bibitem[\protect\citeauthoryear{{Natta}, {Testi}, {Calvet}, {Henning},
  {Waters}  \& {Wilner}}{{Natta} et~al.}{2007}]{Natta:2007ye}
{Natta} A.,  {Testi} L.,  {Calvet} N.,  {Henning} T.,  {Waters} R.,   {Wilner}
  D.,  2007, Protostars and Planets V, \href
  {http://adsabs.harvard.edu/abs/2007prpl.conf..767N} {pp 767--781}

\bibitem[\protect\citeauthoryear{{Okuzumi}, {Momose}, {Sirono}, {Kobayashi}  \&
  {Tanaka}}{{Okuzumi} et~al.}{2016}]{Okuzumi2016}
{Okuzumi} S.,  {Momose} M.,  {Sirono} S.-i.,  {Kobayashi} H.,   {Tanaka} H.,
  2016, \mn@doi [\apj] {10.3847/0004-637X/821/2/82}, \href
  {https://ui.adsabs.harvard.edu/abs/2016ApJ...821...82O} {821, 82}

\bibitem[\protect\citeauthoryear{{Owen}}{{Owen}}{2020}]{Owen2020}
{Owen} J.~E.,  2020, \mn@doi [\mnras] {10.1093/mnras/staa1309}, \href
  {https://ui.adsabs.harvard.edu/abs/2020MNRAS.tmp.1431O} {}

\bibitem[\protect\citeauthoryear{{Paardekooper} \& {Mellema}}{{Paardekooper} \&
  {Mellema}}{2004}]{Paardekooper2004}
{Paardekooper} S.~J.,  {Mellema} G.,  2004, \mn@doi [\aap]
  {10.1051/0004-6361:200400053}, \href
  {https://ui.adsabs.harvard.edu/abs/2004A&A...425L...9P} {425, L9}

\bibitem[\protect\citeauthoryear{{Pani{\'c}}, {Hogerheijde}, {Wilner}  \&
  {Qi}}{{Pani{\'c}} et~al.}{2009}]{Panic2009}
{Pani{\'c}} O.,  {Hogerheijde} M.~R.,  {Wilner} D.,   {Qi} C.,  2009, \mn@doi
  [\aap] {10.1051/0004-6361/200911883}, \href
  {https://ui.adsabs.harvard.edu/abs/2009A&A...501..269P} {501, 269}

\bibitem[\protect\citeauthoryear{{Pegues} et~al.,}{{Pegues}
  et~al.}{2020}]{Pegues2020}
{Pegues} J.,  et~al., 2020, \mn@doi [\apj] {10.3847/1538-4357/ab64d9}, \href
  {https://ui.adsabs.harvard.edu/abs/2020ApJ...890..142P} {890, 142}

\bibitem[\protect\citeauthoryear{{P{\'e}rez} et~al.,}{{P{\'e}rez}
  et~al.}{2012}]{Perez2012}
{P{\'e}rez} L.~M.,  et~al., 2012, \mn@doi [\apjl]
  {10.1088/2041-8205/760/1/L17}, \href
  {https://ui.adsabs.harvard.edu/abs/2012ApJ...760L..17P} {760, L17}

\bibitem[\protect\citeauthoryear{{P{\'e}rez} et~al.,}{{P{\'e}rez}
  et~al.}{2016}]{Perez2016}
{P{\'e}rez} L.~M.,  et~al., 2016, \mn@doi [Science] {10.1126/science.aaf8296},
  \href {https://ui.adsabs.harvard.edu/abs/2016Sci...353.1519P} {353, 1519}

\bibitem[\protect\citeauthoryear{{Pinte} et~al.,}{{Pinte}
  et~al.}{2018}]{Pinte2018planet}
{Pinte} C.,  et~al., 2018, \mn@doi [\apjl] {10.3847/2041-8213/aac6dc}, \href
  {https://ui.adsabs.harvard.edu/abs/2018ApJ...860L..13P} {860, L13}

\bibitem[\protect\citeauthoryear{{Pinte} et~al.,}{{Pinte}
  et~al.}{2019}]{Pinte2019}
{Pinte} C.,  et~al., 2019, \mn@doi [Nature Astronomy]
  {10.1038/s41550-019-0852-6}, \href
  {https://ui.adsabs.harvard.edu/abs/2019NatAs...3.1109P} {3, 1109}

\bibitem[\protect\citeauthoryear{{Pinte} et~al.,}{{Pinte}
  et~al.}{2020}]{Pinte2020}
{Pinte} C.,  et~al., 2020, \mn@doi [\apjl] {10.3847/2041-8213/ab6dda}, \href
  {https://ui.adsabs.harvard.edu/abs/2020ApJ...890L...9P} {890, L9}

\bibitem[\protect\citeauthoryear{{Pollack}, {Hollenbach}, {Beckwith},
  {Simonelli}, {Roush}  \& {Fong}}{{Pollack}
  et~al.}{1994}]{1994ApJ...421..615P}
{Pollack} J.~B.,  {Hollenbach} D.,  {Beckwith} S.,  {Simonelli} D.~P.,  {Roush}
  T.,   {Fong} W.,  1994, \mn@doi [\apj] {10.1086/173677}, \href
  {https://ui.adsabs.harvard.edu/#abs/1994ApJ...421..615P} {421, 615}

\bibitem[\protect\citeauthoryear{{Ricci}, {Testi}, {Natta}, {Neri}, {Cabrit}
  \& {Herczeg}}{{Ricci} et~al.}{2010a}]{Ricci2010a}
{Ricci} L.,  {Testi} L.,  {Natta} A.,  {Neri} R.,  {Cabrit} S.,   {Herczeg}
  G.~J.,  2010a, \mn@doi [\aap] {10.1051/0004-6361/200913403}, \href
  {https://ui.adsabs.harvard.edu/abs/2010A&A...512A..15R} {512, A15}

\bibitem[\protect\citeauthoryear{{Ricci}, {Testi}, {Natta}  \&
  {Brooks}}{{Ricci} et~al.}{2010b}]{Ricci2010b}
{Ricci} L.,  {Testi} L.,  {Natta} A.,   {Brooks} K.~J.,  2010b, \mn@doi [\aap]
  {10.1051/0004-6361/201015039}, \href
  {https://ui.adsabs.harvard.edu/abs/2010A&A...521A..66R} {521, A66}

\bibitem[\protect\citeauthoryear{{Riols} \& {Lesur}}{{Riols} \&
  {Lesur}}{2019}]{Riols2019}
{Riols} A.,  {Lesur} G.,  2019, \mn@doi [\aap] {10.1051/0004-6361/201834813},
  \href {https://ui.adsabs.harvard.edu/abs/2019A&A...625A.108R} {625, A108}

\bibitem[\protect\citeauthoryear{{Rosotti}, {Juhasz}, {Booth}  \&
  {Clarke}}{{Rosotti} et~al.}{2016}]{Rosotti2016}
{Rosotti} G.~P.,  {Juhasz} A.,  {Booth} R.~A.,   {Clarke} C.~J.,  2016, \mn@doi
  [\mnras] {10.1093/mnras/stw691}, \href
  {https://ui.adsabs.harvard.edu/abs/2016MNRAS.459.2790R} {459, 2790}

\bibitem[\protect\citeauthoryear{{Rosotti}, {Tazzari}, {Booth}, {Testi},
  {Lodato}  \& {Clarke}}{{Rosotti} et~al.}{2019}]{Rosotti2019a}
{Rosotti} G.~P.,  {Tazzari} M.,  {Booth} R.~A.,  {Testi} L.,  {Lodato} G.,
  {Clarke} C.,  2019, \mn@doi [\mnras] {10.1093/mnras/stz1190}, \href
  {https://ui.adsabs.harvard.edu/abs/2019MNRAS.486.4829R} {486, 4829}

\bibitem[\protect\citeauthoryear{{Rosotti} et~al.,}{{Rosotti}
  et~al.}{2020a}]{Rosotti2020a}
{Rosotti} G.~P.,  et~al., 2020a, \mn@doi [\mnras] {10.1093/mnras/stz3090},
  \href {https://ui.adsabs.harvard.edu/abs/2020MNRAS.491.1335R} {491, 1335}

\bibitem[\protect\citeauthoryear{{Rosotti}, {Teague}, {Dullemond}, {Booth}  \&
  {Clarke}}{{Rosotti} et~al.}{2020b}]{Rosotti2020b}
{Rosotti} G.~P.,  {Teague} R.,  {Dullemond} C.,  {Booth} R.~A.,   {Clarke}
  C.~J.,  2020b, \mn@doi [\mnras] {10.1093/mnras/staa1170}, \href
  {https://ui.adsabs.harvard.edu/abs/2020MNRAS.495..173R} {495, 173}

\bibitem[\protect\citeauthoryear{{Sch{\"o}ier}, {van der Tak}, {van Dishoeck}
  \& {Black}}{{Sch{\"o}ier} et~al.}{2005}]{Schoeier2005}
{Sch{\"o}ier} F.~L.,  {van der Tak} F.~F.~S.,  {van Dishoeck} E.~F.,   {Black}
  J.~H.,  2005, \mn@doi [\aap] {10.1051/0004-6361:20041729}, \href
  {https://ui.adsabs.harvard.edu/abs/2005A&A...432..369S} {432, 369}

\bibitem[\protect\citeauthoryear{{Semenov} et~al.,}{{Semenov}
  et~al.}{2018}]{Semenov2018}
{Semenov} D.,  et~al., 2018, \mn@doi [\aap] {10.1051/0004-6361/201832980},
  \href {https://ui.adsabs.harvard.edu/abs/2018A&A...617A..28S} {617, A28}

\bibitem[\protect\citeauthoryear{{Suriano}, {Li}, {Krasnopolsky}  \&
  {Shang}}{{Suriano} et~al.}{2018}]{Suriano2018}
{Suriano} S.~S.,  {Li} Z.-Y.,  {Krasnopolsky} R.,   {Shang} H.,  2018, \mn@doi
  [\mnras] {10.1093/mnras/sty717}, \href
  {https://ui.adsabs.harvard.edu/abs/2018MNRAS.477.1239S} {477, 1239}

\bibitem[\protect\citeauthoryear{{Tazzari} et~al.,}{{Tazzari}
  et~al.}{2016}]{Tazzari2016}
{Tazzari} M.,  et~al., 2016, \mn@doi [\aap] {10.1051/0004-6361/201527423},
  \href {https://ui.adsabs.harvard.edu/abs/2016A&A...588A..53T} {588, A53}

\bibitem[\protect\citeauthoryear{{Tazzari}, {Beaujean}  \& {Testi}}{{Tazzari}
  et~al.}{2018}]{GalarioPaper}
{Tazzari} M.,  {Beaujean} F.,   {Testi} L.,  2018, \mn@doi [\mnras]
  {10.1093/mnras/sty409}, \href
  {https://ui.adsabs.harvard.edu/\#abs/2018MNRAS.476.4527T} {476, 4527}

\bibitem[\protect\citeauthoryear{{Tazzari}, {Clarke}, {Testi}, {Williams},
  {Facchini}, {Manara}, {Natta}  \& {Rosotti}}{{Tazzari}
  et~al.}{2020a}]{Tazzari2020}
{Tazzari} M.,  {Clarke} C.~J.,  {Testi} L.,  {Williams} J.~P.,  {Facchini} S.,
  {Manara} C.~F.,  {Natta} A.,   {Rosotti} G.,  2020a, arXiv e-prints, \href
  {https://ui.adsabs.harvard.edu/abs/2020arXiv201002249T} {p. arXiv:2010.02249}

\bibitem[\protect\citeauthoryear{{Tazzari}, {Clarke}, {Testi}, {Williams},
  {Facchini}, {Manara}, {Natta}  \& {Rosotti}}{{Tazzari}
  et~al.}{2020b}]{2020arXiv201002249T}
{Tazzari} M.,  {Clarke} C.~J.,  {Testi} L.,  {Williams} J.~P.,  {Facchini} S.,
  {Manara} C.~F.,  {Natta} A.,   {Rosotti} G.,  2020b, arXiv e-prints, \href
  {https://ui.adsabs.harvard.edu/abs/2020arXiv201002249T} {p. arXiv:2010.02249}

\bibitem[\protect\citeauthoryear{{Teague}}{{Teague}}{2019a}]{2019JOSS....4.1220T}
{Teague} R.,  2019a, \mn@doi [The Journal of Open Source Software]
  {10.21105/joss.01220}, \href
  {https://ui.adsabs.harvard.edu/abs/2019JOSS....4.1220T} {4, 1220}

\bibitem[\protect\citeauthoryear{Teague}{Teague}{2019b}]{teague_gf_2019}
Teague R.,  2019b, \mn@doi [The Journal of Open Source Software]
  {10.21105/joss.01632}, 4, 1632

\bibitem[\protect\citeauthoryear{{Teague} \& {Foreman-Mackey}}{{Teague} \&
  {Foreman-Mackey}}{2018}]{teague_bm_2018}
{Teague} R.,  {Foreman-Mackey} D.,  2018, \mn@doi [Research Notes of the
  American Astronomical Society] {10.3847/2515-5172/aae265}, \href
  {https://ui.adsabs.harvard.edu/abs/2018RNAAS...2c.173T} {2, 173}

\bibitem[\protect\citeauthoryear{{Teague} et~al.,}{{Teague}
  et~al.}{2017}]{Teague2017}
{Teague} R.,  et~al., 2017, \mn@doi [\apj] {10.3847/1538-4357/835/2/228}, \href
  {https://ui.adsabs.harvard.edu/abs/2017ApJ...835..228T} {835, 228}

\bibitem[\protect\citeauthoryear{{Teague}, {Bae}, {Bergin}, {Birnstiel}  \&
  {Foreman-Mackey}}{{Teague} et~al.}{2018}]{Teague2018}
{Teague} R.,  {Bae} J.,  {Bergin} E.~A.,  {Birnstiel} T.,   {Foreman-Mackey}
  D.,  2018, \mn@doi [\apjl] {10.3847/2041-8213/aac6d7}, \href
  {https://ui.adsabs.harvard.edu/abs/2018ApJ...860L..12T} {860, L12}

\bibitem[\protect\citeauthoryear{{Testi} et~al.,}{{Testi}
  et~al.}{2014}]{Testi2014}
{Testi} L.,  et~al., 2014, in {Beuther} H.,  {Klessen} R.~S.,  {Dullemond}
  C.~P.,   {Henning} T.,  eds, Protostars and Planets VI. p.~339 (\mn@eprint
  {arXiv} {1402.1354}), \mn@doi{10.2458/azu_uapress_9780816531240-ch015}

\bibitem[\protect\citeauthoryear{{Trapman}, {Facchini}, {Hogerheijde}, {van
  Dishoeck}  \& {Bruderer}}{{Trapman} et~al.}{2019}]{Trapman2019}
{Trapman} L.,  {Facchini} S.,  {Hogerheijde} M.~R.,  {van Dishoeck} E.~F.,
  {Bruderer} S.,  2019, \mn@doi [\aap] {10.1051/0004-6361/201834723}, \href
  {https://ui.adsabs.harvard.edu/abs/2019A&A...629A..79T} {629, A79}

\bibitem[\protect\citeauthoryear{{Tsukagoshi} et~al.,}{{Tsukagoshi}
  et~al.}{2016}]{Tsukagoshi2016}
{Tsukagoshi} T.,  et~al., 2016, \mn@doi [\apjl] {10.3847/2041-8205/829/2/L35},
  \href {https://ui.adsabs.harvard.edu/abs/2016ApJ...829L..35T} {829, L35}

\bibitem[\protect\citeauthoryear{{Visser}, {van Dishoeck}  \& {Black}}{{Visser}
  et~al.}{2009}]{Visser2009}
{Visser} R.,  {van Dishoeck} E.~F.,   {Black} J.~H.,  2009, \mn@doi [\aap]
  {10.1051/0004-6361/200912129}, \href
  {https://ui.adsabs.harvard.edu/abs/2009A&A...503..323V} {503, 323}

\bibitem[\protect\citeauthoryear{{Weaver}, {Isella}  \& {Boehler}}{{Weaver}
  et~al.}{2018}]{Weaver2018}
{Weaver} E.,  {Isella} A.,   {Boehler} Y.,  2018, \mn@doi [\apj]
  {10.3847/1538-4357/aaa481}, \href
  {https://ui.adsabs.harvard.edu/abs/2018ApJ...853..113W} {853, 113}

\bibitem[\protect\citeauthoryear{{Williams} \& {Best}}{{Williams} \&
  {Best}}{2014}]{WilliamsBest2014}
{Williams} J.~P.,  {Best} W. M.~J.,  2014, \mn@doi [\apj]
  {10.1088/0004-637X/788/1/59}, \href
  {https://ui.adsabs.harvard.edu/abs/2014ApJ...788...59W} {788, 59}

\bibitem[\protect\citeauthoryear{{Youdin} \& {Goodman}}{{Youdin} \&
  {Goodman}}{2005}]{YoudinGoodman2005}
{Youdin} A.~N.,  {Goodman} J.,  2005, \mn@doi [\apj] {10.1086/426895}, \href
  {https://ui.adsabs.harvard.edu/abs/2005ApJ...620..459Y} {620, 459}

\bibitem[\protect\citeauthoryear{{Zhang}, {Blake}  \& {Bergin}}{{Zhang}
  et~al.}{2015}]{Zhang2015}
{Zhang} K.,  {Blake} G.~A.,   {Bergin} E.~A.,  2015, \mn@doi [\apjl]
  {10.1088/2041-8205/806/1/L7}, \href
  {https://ui.adsabs.harvard.edu/abs/2015ApJ...806L...7Z} {806, L7}

\bibitem[\protect\citeauthoryear{{Zhang} et~al.,}{{Zhang}
  et~al.}{2018}]{Zhang2018}
{Zhang} S.,  et~al., 2018, \mn@doi [\apjl] {10.3847/2041-8213/aaf744}, \href
  {https://ui.adsabs.harvard.edu/abs/2018ApJ...869L..47Z} {869, L47}

\bibitem[\protect\citeauthoryear{{van Dishoeck} \& {Black}}{{van Dishoeck} \&
  {Black}}{1988}]{vanDishoeckBlack1988}
{van Dishoeck} E.~F.,  {Black} J.~H.,  1988, \mn@doi [\apj] {10.1086/166877},
  \href {https://ui.adsabs.harvard.edu/abs/1988ApJ...334..771V} {334, 771}

\bibitem[\protect\citeauthoryear{{van der Marel} et~al.,}{{van der Marel}
  et~al.}{2013}]{vanderMarel2013}
{van der Marel} N.,  et~al., 2013, \mn@doi [Science] {10.1126/science.1236770},
  \href {https://ui.adsabs.harvard.edu/abs/2013Sci...340.1199V} {340, 1199}

\makeatother
\end{thebibliography}



\appendix



\section{The $^{12}$CO emission}

\begin{figure*}
\centering
\includegraphics[width=\textwidth]{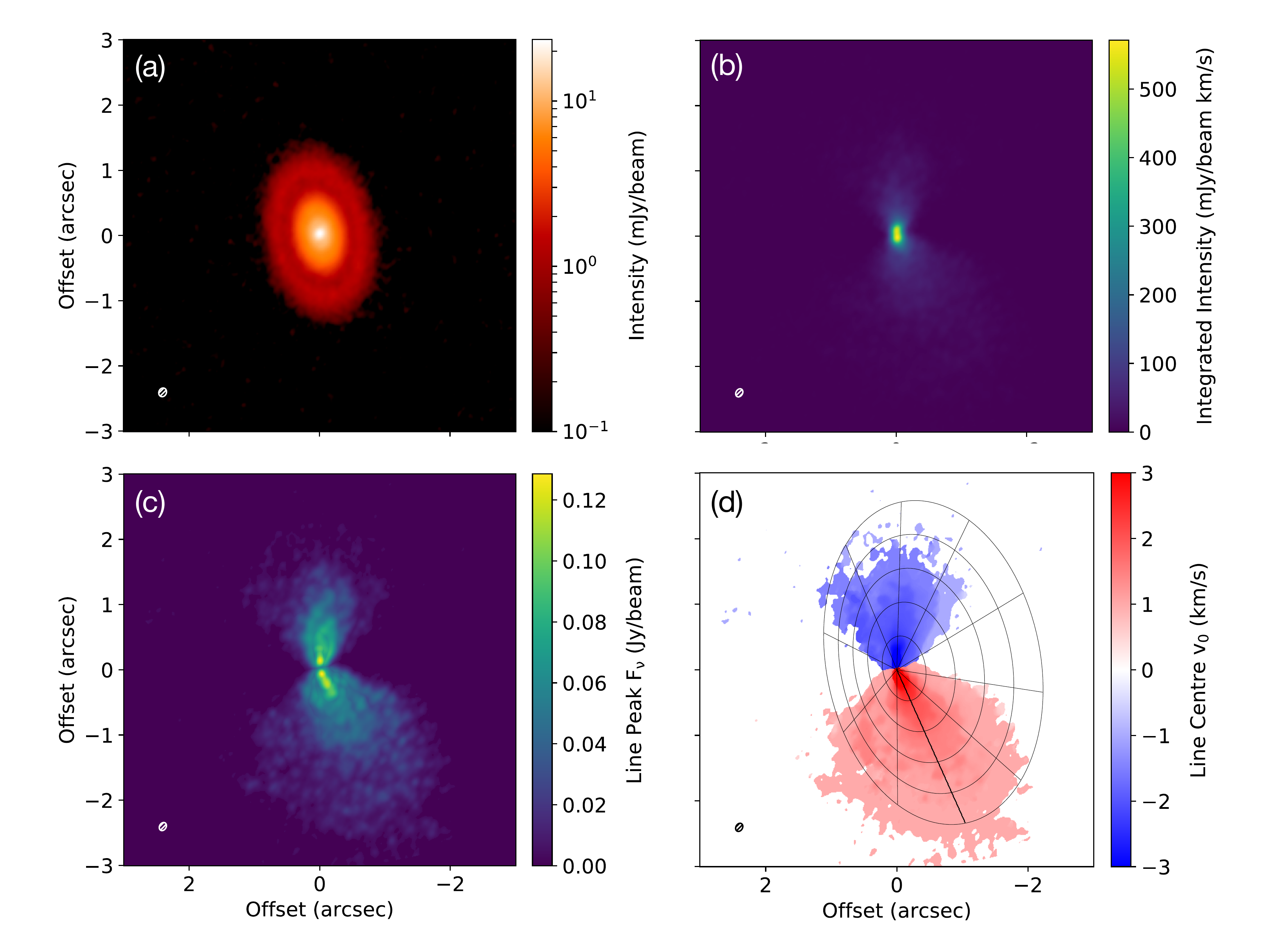}
\caption{As \autoref{fig:13CO_observations}, but for $^{12}$CO (3--2). Channels approximately $\pm1.5$ km\,s$^{-1}$ from the line centre are affected by cloud contamination.  The line centre panel is overlaid with a conical surface of aspect ratio $z/r=0.3$.}
\label{fig:12CO_observations}
\end{figure*}

Figure \autoref{fig:12CO_observations} outlines the $^{12}$CO observations.

\section{On continuum subtraction}
\label{append:continuum_subtraction}

\begin{figure}
\includegraphics[width=\columnwidth]{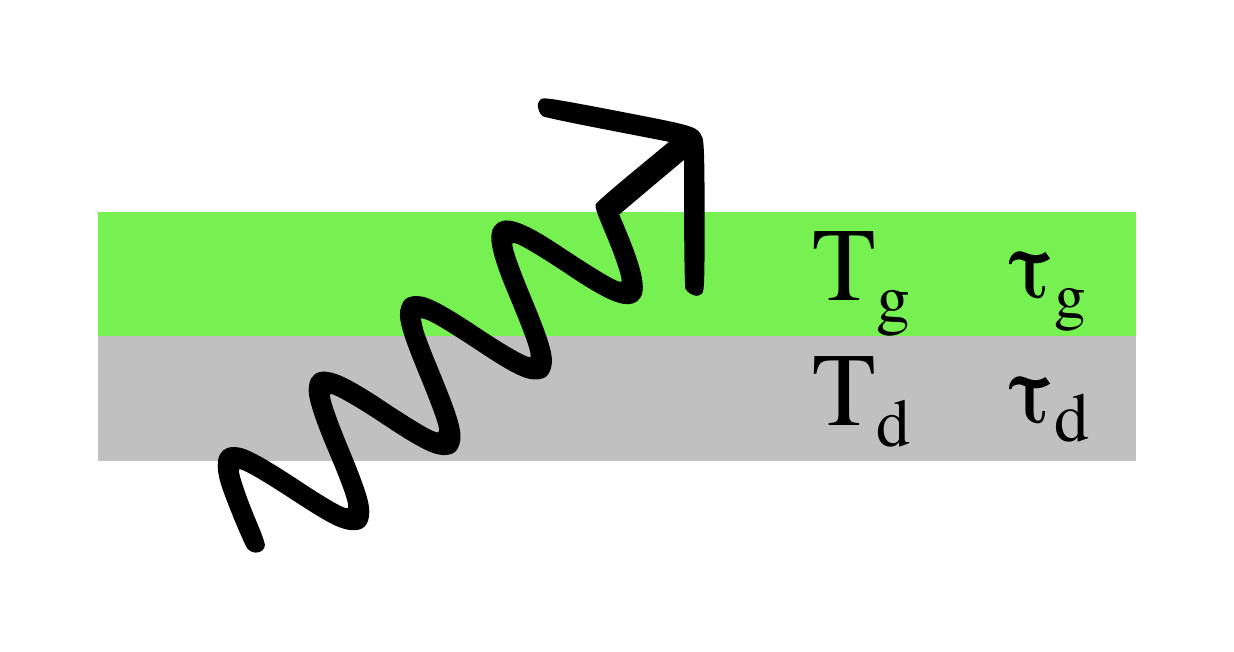}
\caption{Graphical depiction of the simplified model we use to illustrate why continuum subtraction is not necessary. We assume there is a gas layer with temperature $T_g$ and optical depth $\tau_g$ in front of a dust layer with temperature $T_d$ and optical depth $\tau_d$. The total emission is given by \autoref{eq:intensity}.}
\label{fig:schematics_layers}
\end{figure}

\subsection{Intensity in a channel map} 

To understand why continuum subtraction can create spurious structure in gas maps, we start writing the emission $I_\nu$ at a given location in the image, and at a given frequency $\nu$, as:
\begin{equation}
I_\nu=[1-\exp(-\tau_g)]B_\nu(T_g) + [1-\exp(-\tau_d)]\exp(-\tau_g) B_\nu(T_d),
\label{eq:intensity}
\end{equation}
where $\tau$ denotes optical depth, $T$ temperature, and $B_\nu$ is the Planck function, and we have used the suffix $_g$ for the gas and $_d$ for the dust. This equation assumes the situation depicted in \autoref{fig:schematics_layers}, i.e. a dust layer behind a gas layer; each layer emits radiation and the gas layer also partly absorbs the dust continuum emission. This assumption is justified for molecules like CO that are abundant in the warm molecular layer, while the dust is typically settled to the midplane. 

The goal of performing continuum subtraction is to recover the intrinsic line emission, that is, only the first term $[1-\exp(-\tau_g)]B_\nu(T_g)$. Continuum subtraction is justified if line emission is optically thin, because in this case the attenuation of the dust from the gas is negligible, i.e. $\exp(-\tau_g) \simeq 1$. We note however that subtraction is not necessary in the case line emission is optically thick, because in this case the second term is completely absorbed by the gas in front. Subtracting the continuum in this case is not only not necessary, but erroneous because it leads to an underestimate of the intrinsic gas emission. The brighter the continuum is with respect to the gas, the more severe the underestimate is. This explains why continuum subtraction can lead to a spurious ``gap'' in the gas emission at a location where the continuum is bright, i.e. it has a ``ring'', as we show in section \ref{ssec:cont_sub}. Note that the creation of spurious sub-structure does not require that the continuum is optically thick, although it does require that the surface brightness of the continuum is comparable to the line emission.

Of course, the true structure of proto-planetary discs is more complex than captured in \autoref{eq:intensity}, and it is not true that the CO gas has a single temperature. However, the fact that we should not subtract the continuum if the line emission is optically thick still stands. We also note that in principle we would need to add another gas layer, placed behind the dust, to account for the fact that the backside of the disc also contributes to the emission. However, if in a given channel the front layer is optically thick, the back layer is completely absorbed and does not affect this discussion.

\subsection{Intensity in a moment map} 

Up to now we have considered a single channel. In reality, due to Keplerian rotation, even if emission is optically thick in a given channel, it will become optically thin for channels sufficiently far in velocity space. When constructing an integrated intensity map, this implies that understanding whether to apply continuum subtraction becomes extremely complex: at a given location some channels are optically thick (and therefore should not be continuum subtracted) whereas other ones are optically thin (and should be continuum subtracted). Stated in another way, to recover the intrinsic gas emission, we should know the optical depth $\tau_g$ of the gas in each channel.

It is conceptually easier instead to compute a peak intensity map. In this case, provided at least one channel is optically thick, the peak brightness will simply be the intrinsic line emission, unaffected by the continuum, and continuum subtraction is therefore not necessary. The example of section \ref{ssec:cont_sub} confirms that this intuition is correct.

In concluding, we note that this discussion is largely similar to \citet{Weaver2018} (see also \citealt{Boehler2017}). In that case, the authors focused on correctly measuring the gas temperature rather than the intrinsic gas emission, but we note that for optically thick emission the two things are equivalent. The important difference is that \citet{Weaver2018} only presented models of smooth discs. Instead, what we show in this section, in the example of section \ref{ssec:cont_sub} and on the observational data, is that, when dealing with discs showing sub-structure, continuum subtraction can lead to the creation of artificial sub-structures in the gas emission, such as dark gaps at the location of bright continuum rings.

\section{Height of the emission surface}

\begin{figure}
\includegraphics[width=\columnwidth]{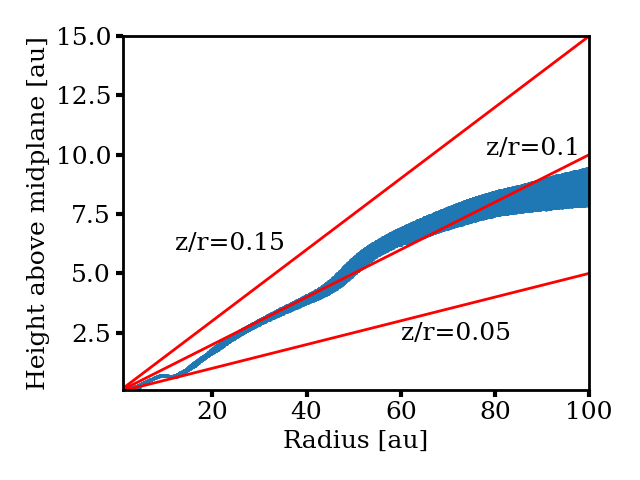}
\caption{Height of the emission surface for the ``constant dust-to-gas ratio'' model.}
\label{fig:emission_surface}
\end{figure}

Figure \autoref{fig:emission_surface} shows the height of the emission surface in the ``constant dust-to-gas ratio'' model.

\section{Channel maps}
\label{sec:chans}
\begin{figure*}
\centering
\includegraphics[width=0.95\textwidth,trim={1.5cm 1.0cm 0 1.25cm},clip]{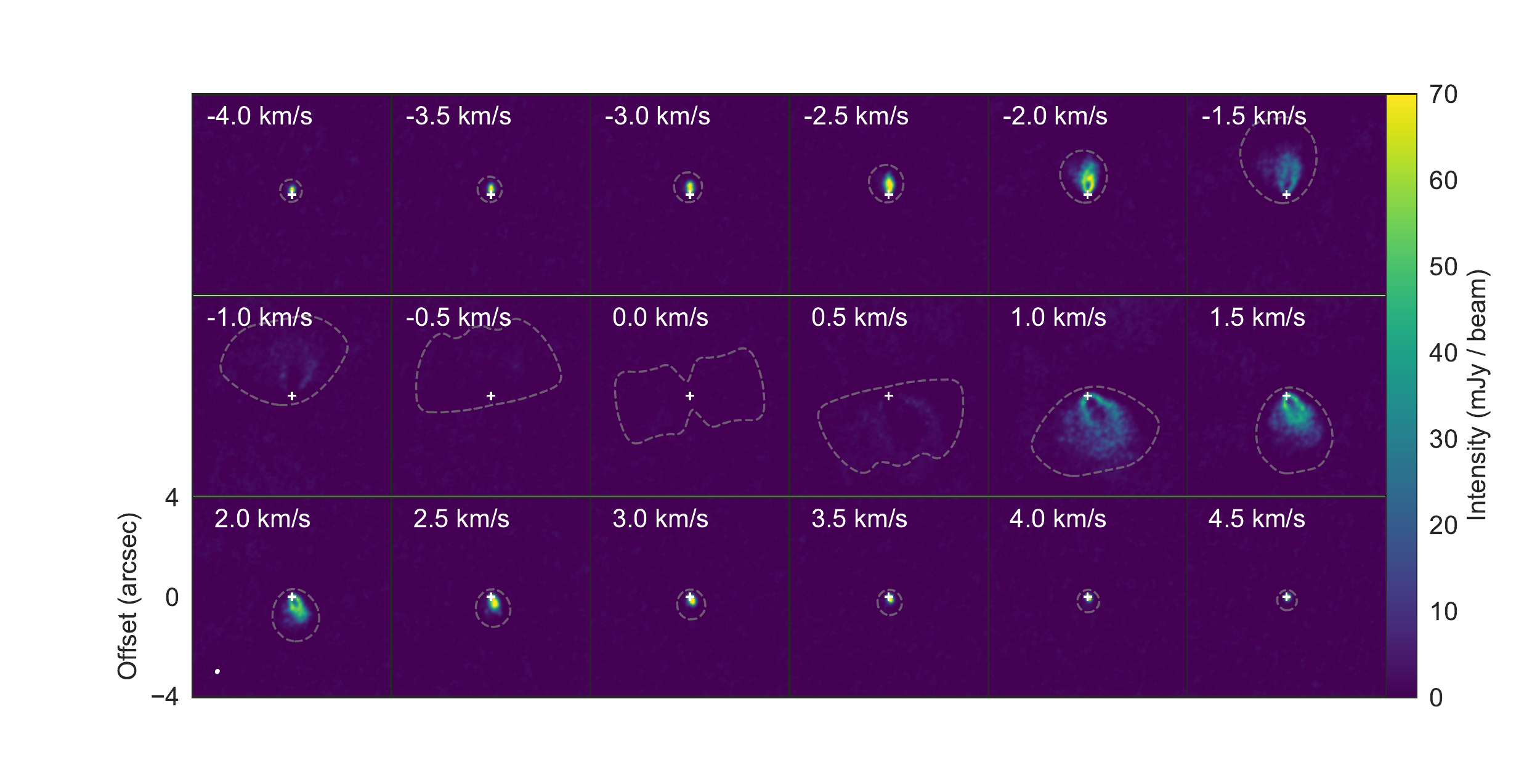}
\includegraphics[width=0.95\textwidth,trim={1.5cm 1.0cm 0 1.25cm},clip]{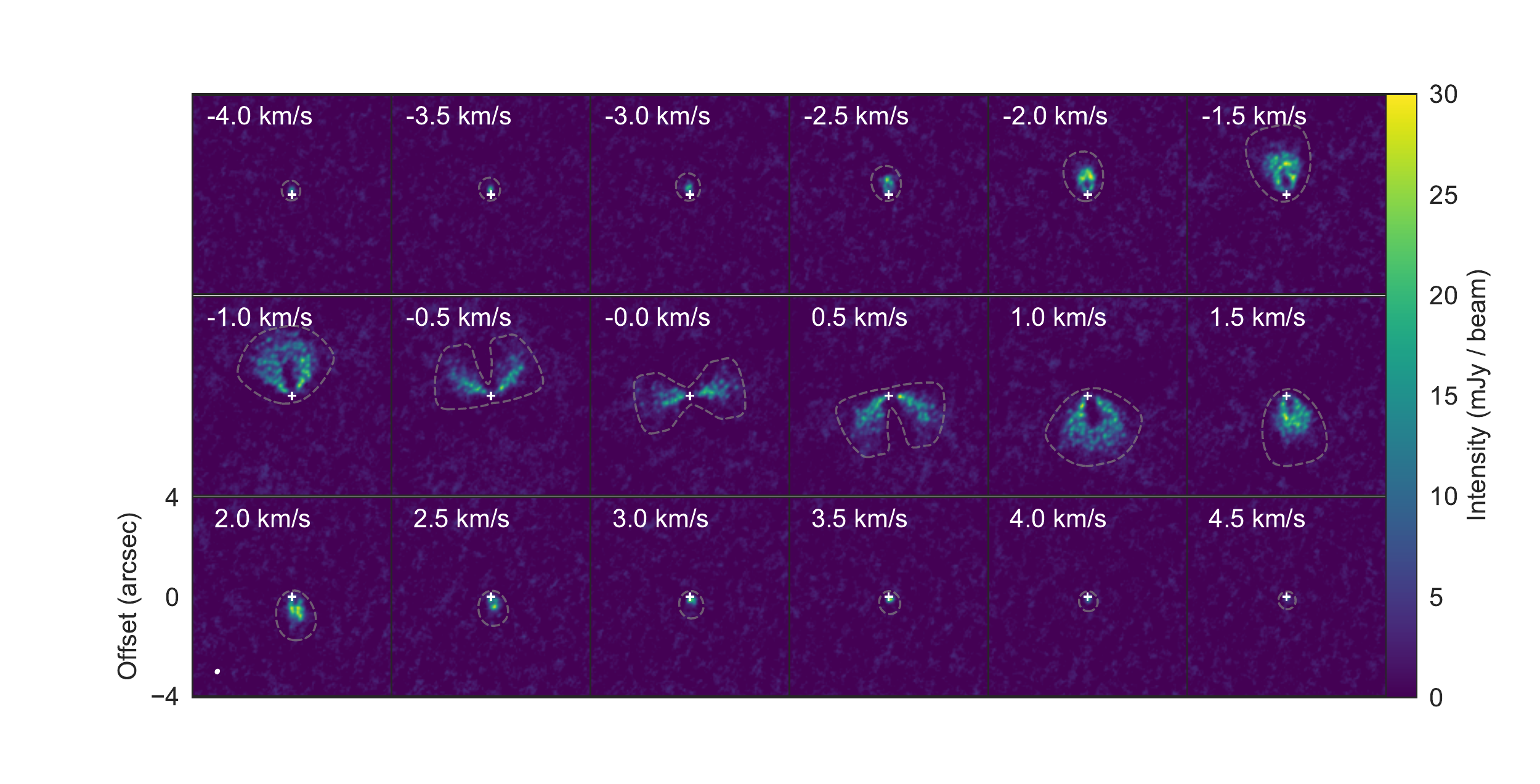}
\includegraphics[width=0.95\textwidth,trim={1.5cm 0.0cm 0 1.25cm},clip]{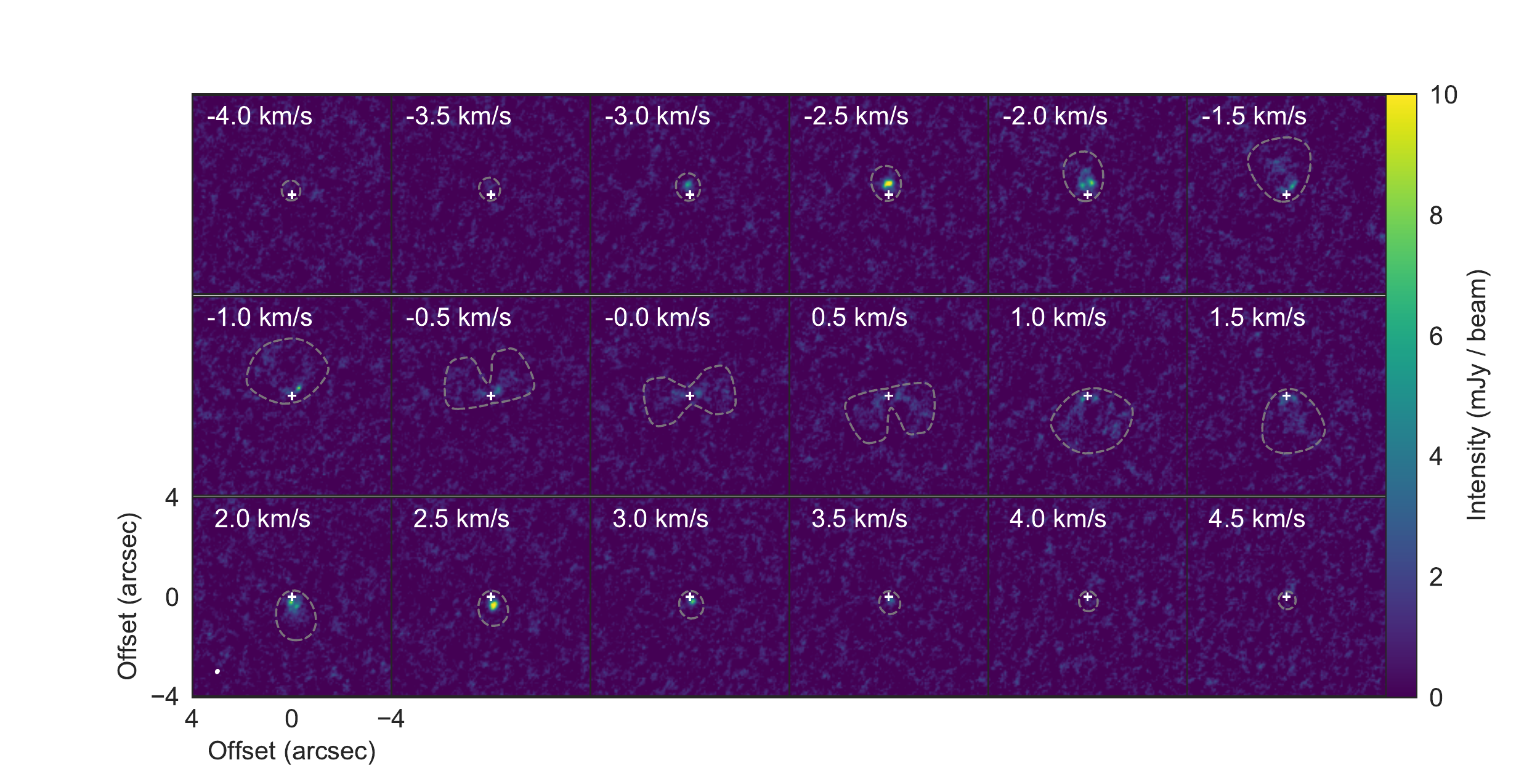}
\caption{Channel maps for $^{12}$CO (top), $^{13}$CO (middle) and CS (bottom).  Keplerian masks used during imaging and moment map generation are shown with a dashed line.}
\label{fig:channel_maps}
\end{figure*}

Figure \ref{fig:channel_maps} shows the channel maps for $^{12}$CO, $^{13}$CO and CS.

\bsp	
\label{lastpage}
\end{document}